\documentclass[fleqn,usenatbib]{mnras}

\usepackage{newtxtext,newtxmath}

\usepackage[T1]{fontenc}

\DeclareRobustCommand{\VAN}[3]{#2}
\let\VANthebibliography\thebibliography
\def\thebibliography{\DeclareRobustCommand{\VAN}[3]{##3}\VANthebibliography}

\usepackage{graphicx}	
\usepackage{amsmath}	

\usepackage{enumitem}

\title[Galaxy clumps at Cosmic Noon]{Clumps as multiscale structures in Cosmic Noon galaxies}

\author[B. S. Kalita et al.]{Boris S. Kalita,$^{1,2,3}$\thanks{E-mail: boris.kalita@ipmu.jp; kalita.boris.sindhu@gmail.com}\thanks{Joint-Kavli Astrophysics Fellow}
Tomoko L.  Suzuki,$^{1,3}$
Daichi Kashino,$^{4}$
John D. Silverman,$^{1,5,6,3}$
Emanuele Daddi,$^{7}$
\newauthor 
Luis C. Ho,$^{2,8}$
Xuheng Ding,$^{9}$
Wilfried Mercier,$^{10}$
Andreas L. Faisst,$^{11}$
Kartik Sheth,$^{12}$
Francesco Valentino,$^{13.14}$
\newauthor 
Annagrazia Puglisi,$^{15}$
Toshiki Saito,$^{16}$
Darshan Kakkad,$^{17}$
Olivier Ilbert,$^{10}$
Ali Ahmad Khostovan,$^{18,19}$
\newauthor
Zhaoxuan Liu,$^{1,3,20}$
Takumi Tanaka,$^{1,3,20}$
Georgios Magdis,$^{13,21,14}$
Jorge A. Zavala,$^{4}$
Qinghua Tan,$^{22}$
\newauthor
Jeyhan S. Kartaltepe,$^{18}$ 
Lilan Yang,$^{18}$
Anton M. Koekemoer,$^{23}$
Jed McKinney,$^{24}$
Brant E. Robertson,$^{25}$
\newauthor
Shuowen Jin,$^{13,21}$
Christopher C. Hayward,$^{26}$
Michaela Hirschmann,$^{27,28}$
Maximilien Franco,$^{24}$
\newauthor
Marko Shuntov,$^{13,14}$
Ghassem Gozaliasl,$^{29,30}$
Aidan Kaminsky,$^{31}$
and R. Michael Rich$^{32}$
\\ \\
Affiliations are listed at the end of the paper
}

\date{Accepted 2024 December 17. Received 2024 December 6; in original form 2024 September 20}

\pubyear{\the\year{}}

\begin{document}
\label{firstpage}
\pagerange{\pageref{firstpage}--\pageref{lastpage}}
\maketitle

\begin{abstract}
Star-forming clumps have been found to significantly influence the star formation of gas-rich $z>1$ galaxies. Using public data from JWST/NIRCam (Cosmic Evolution Survey; COSMOS-Web) and Atacama Large (sub-)Millimeter Array (ALMA; Fiber-Multi Object Spectrograph or FMOS-COSMOS survey), we study a sample of 32 massive ($>10^{10.5}\,\rm M_{\odot}$) main-sequence galaxies at $z_{\rm spec}\sim 1.5$ with $\sim 0.3\,\rm kpc$ resolution. We create composite morphological models consisting of bulge, disk, and clumps to fully `deconstruct’ the galaxy images. With the resulting measurements of the flux and size of these components, we find the following: (I) The combined contribution of clumps is $1-30\%$ towards the net star formation of the host while contributing $1-20\%$ to its stellar mass.  The clumps show a correlation between their stellar mass and star formation rate (SFR), but have an increased specific SFR relative to the star-forming main-sequence,  with offsets ranging from $0 \lesssim \Delta \log \rm sSFR \lesssim 0.4$. They feature star-formation surface densities of $10^{-2}-10^{2}\,\rm M_{\odot}/yr/kpc^{2}$, consistent with values observed in both local star-forming and starburst galaxies. (II) The detected clumps span a large range of characteristic sizes ($r_{e} \sim 0.1 - 1\,\rm kpc$) and stellar masses ($\sim 10^{8.0-9.5}\,\rm M_{\odot}$). We estimate a mass-size relation ($r_{e} \propto \rm M_{\star}^{\,0.52 \pm 0.07}$) along with a stellar mass function (slope, $\alpha = - 1.85 \pm 0.19$),  both suggesting a hierarchical nature similar to that expected in star-forming regions in local galaxies. (III) Our measurements agree with the properties of stellar clumps in $z\gtrsim1$ lensed systems, bridging the gap between lensed and unlensed studies by detecting structures at sub-kpc scales. (IV) Clumps are found to be preferentially located along spiral features visible primarily in the residual rest-frame near-IR images. In conclusion, we present an observation-based, coherent picture of star-forming clumps in galaxies at $z > 1$.
\end{abstract}



\section{Introduction} \label{sec:jwst_clumps}
The highly clumpy nature of star-forming galaxies at $z>1$ has been recognized for nearly three decades \citep{abraham96, bergh96,elmegreen05,elmegreen08,forster11, wuyts12, guo15,guo18,shibuya16, soto17, huertas-company20}. Due to their high gas fractions, these galaxies likely experience instabilities from fragmentation \citep{daddi2008, daddi10, tacconi10, tacconi13, geach11, magdis12, bethermin15,rujopakarn23,claeyssens23}.  Stellar clumps (which will be referred to as `clumps' throughout this work) formed under such conditions are expected to be more massive than similar structures observed at lower redshifts \citep{cowie95,cook16,larson20,mehta21}.  The larger sizes likely reflect the greater characteristic lengths of disk instabilities at higher redshifts \citep{elmegreen08,genzel11,genzel23}. However, details of clump properties remain debated. The lack of clarity primarily stems from disagreements about the true distribution of mass, size, and ages of these structures, with observations of lensed and unlensed galaxies, as well as simulations, differing in their conclusions.

A defining feature of clumps that is agreed upon however,  is their significant contribution to the total star formation of the host galaxy. In the local universe, stars typically form within giant molecular clouds \citep[][for reviews]{mckee07,schinnerer24}. However, studies of star-forming galaxies at $z>1$ indicate the formation of giant star-forming regions due to dynamical instabilities on scales approaching the disk thickness \citep{elmegreen09, jones10, livermore12, bournaud16,genzel23}. This is further supported by measurements showing the Toomre instability parameter is $<1$ in regions where clumps are detected \citep{genzel11}.

Theoretically, clumps arising from instabilities will have their mass and size as a function of the surface density of the host disks \citep{livermore12, ceverino12,reina-campos17}. The expected (and observed) sizes of these clumps can range from $\sim 100\,\rm pc$ \citep[or lower, e.g.,][similar to local massive star-forming HII regions]{cava18} to $\sim 1\,\rm kpc$ \citep{elmegreen09, jones10}, with the massive end requiring higher dispersion or rotation to counterbalance increased self-gravity \citep{dekel06, epinat12, ceverino12,livermore15}. 

The spatial resolution likely limits detection capabilities, with lensed studies detecting scales of a few hundred to tens of parsecs \citep{jones10, livermore12, adamo13, johnson17,cava18, zick20,vanzella22a,vanzella22b,mestric22,messa22,claeyssens23,adamo24,fujimoto24,mowla24,messa24,claeyssens24}, but potentially resolving out more extended structures (depending on the signal-to-noise ratio after magnification). Meanwhile, un-lensed studies, primarily using HST, are affected by blending,  and mainly detect clumps $\gtrsim 1\,\rm kpc$ \citep[e.g.,][]{genzel06, forster11, guo18}. Such resolution limits can lead to an overestimation of stellar masses and sizes of the clumps \citep{tamburello15, fisher17, huertas-company20}.

\begin{figure*} 
    \centering
    \includegraphics[width=\textwidth]{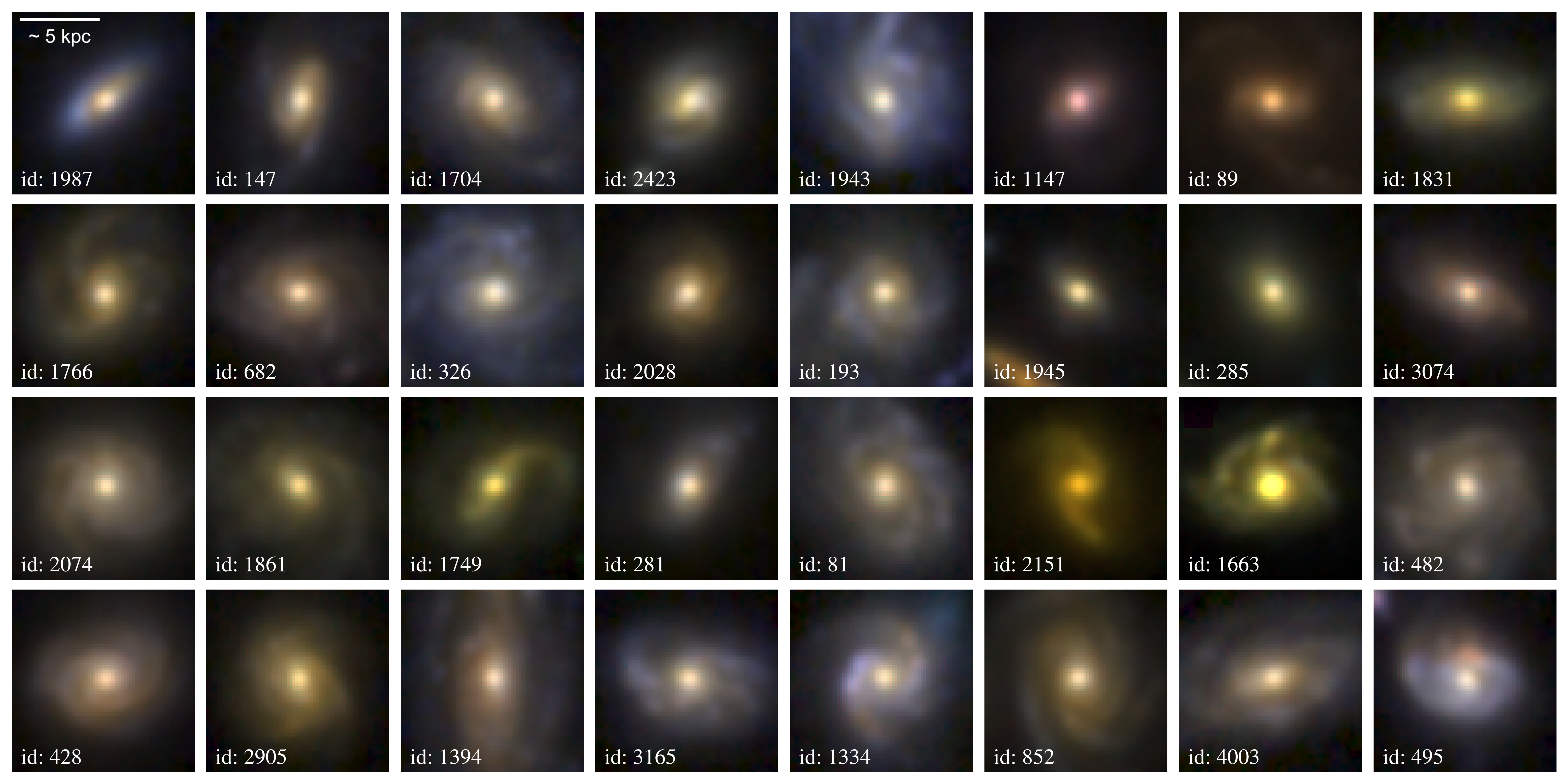}
    \caption{The RGB images (F150W, F277W, F444W) of the 32 galaxies used in this work have dimensions of $70 \times 70~$pixels, or $2.1^{\prime\prime} \times 2.1^{\prime\prime}$. They are arranged in order of increasing RA and Dec. All are within a redshift range of $1.43 \leq z \leq 1.74$ and have stellar masses of $10^{10.5-11.4}\,\rm M_{\odot}$. The FMOS ID for each galaxy is provided in the bottom-left corner, with their corresponding properties listed in Table~\ref{tab:1} in the Appendix. The F150W (blue) and F277W (green) images are PSF-matched to the F444W (red) image using a Gaussian kernel. However, this work does not require PSF matching; it is performed here solely for illustration purposes.}
    \label{fig:image_comp}
\end{figure*}

The implication of accurate measurements goes beyond the characterisation of clumps and the conditions of star formation. There is a debate regarding the survival of these structures, with some simulations suggesting that they survive up to a few $100\,\rm Myrs$. This inevitably leads to an inward migration into the forming bulge due to the combined effect of dynamical friction and gravitational torques \citep{elmegreen07, mandelker14, mandelker17}. Although, massive clumps could be sheered by the gravitational potential well if they grow larger than the Jeans length \citep{elmegreen96}. On the other hand,  some works suggest that they are disrupted due to stellar feedback on very short timescales $\lesssim 100\,\rm Myr$ \citep{murray10, hopkins12, hopkins14, buck17, oklopvic17} and play an insignificant role in bulge formation.  In either case, the bulge once formed can stabilise the disk and lead to a decrease in the formation of clumps \citep{martig09,hopkins23, kalita24}. 

This uncertainty could be resolved with proper measurements of clump masses and sizes across the aforementioned variety of scales. \cite{mandelker17} found that clumps with masses $>10^{8.2}\,\rm M_{\odot}$ would survive the effects of feedback, while those that are smaller would be short-lived. \cite{krumholz10} provide a division based on star formation. Clumps would be disrupted if $>10\%$ of their mass is converted to stars within their free-fall time.

Recent works targeting clumps in  un-lensed galaxies at $z>1$ have begun using the JWST/NIRCam (James Webb Space Telescope/Near-IR camera) instrument \citep{kalita24, kalita24b}. In these studies,  the wide wavelength coverage is used to measure aperture-based and background subtracted clump fluxes. This however restricts the resolution to the longest wavelength band due to the need for PSF-matching (F444W; with resolution of $\gtrsim 1\,\rm kpc$ at $z>1.0$). Hence the higher resolutions of the shorter wavelength bands (F115W and F150W) are not fully exploited.

Use of F115W and F150W bands would lead to point spread function (PSF) full width half maximum (FWHM) $\lesssim 400\,\rm pc$ at $z\sim1.5$.  If clumps were detected at $3 \sigma$, one can therefore have spatial sensitivity down to about $0.1\,\rm kpc$ for the F150W filter at the target redshift of this work ($z \approx 1.5$). Therefore, physical resolutions usually achievable in lensed studies before jWST can now be naturally reached.  In this work, the clump measurements are done using model fitting rather than apertures. This allows the extraction of fluxes through de-blending even in the longer JWST bands with slightly broader PSFs . This method also allows us to reach spatial sensitivities down to $0.1\,\rm kpc$ (which is empirically verified in Sec.~\ref{subsec:completeness}). 

Throughout this paper, we adopt a concordance $\Lambda$CDM cosmology, characterized by  $\Omega_{\rm m}=0.3$, $\Omega_{\Lambda}=0.7$, and $\rm H_{0}=70~\rm km\, s^{-1}\,\rm Mpc^{-1}$. Magnitudes and colors are on the AB scale.  All images are oriented such that north is up and east is left.  Use of the $\log$-scale always refers to $\log_{10}$. 
\begin{figure*} 
    \centering
    \includegraphics[width=\textwidth]{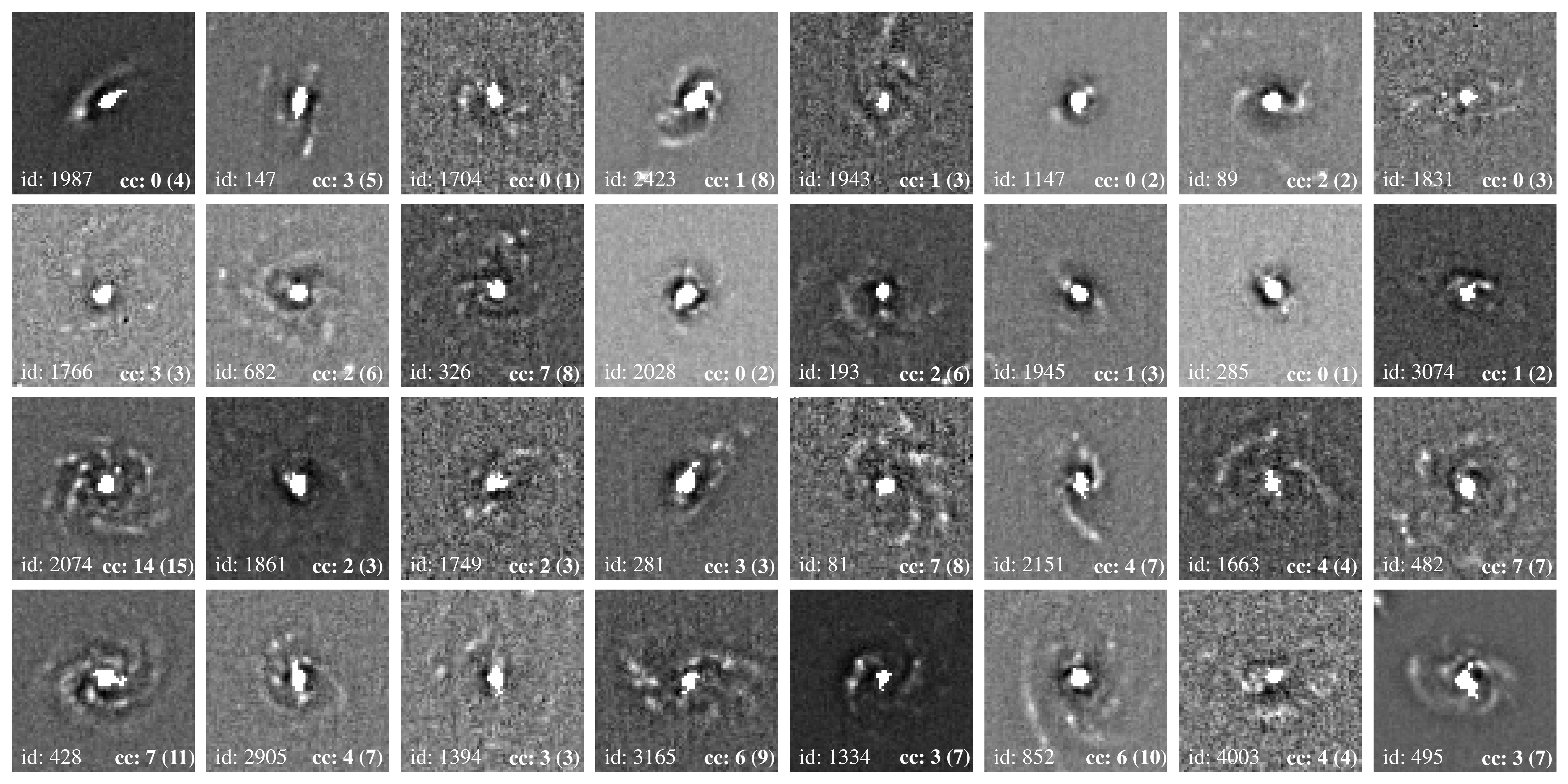}
    \caption{The contrast images created from the F150W images, by subtracting a smoothed version of themselves, for the 32 galaxies in our sample (RGB images shown in Fig.~\ref{fig:image_comp}). These are used to detect clumps.  In each galaxy, we have also added the final number of detected clumps (clump counts,  or cc),  along with initial number of clumps, shown in brackets, before the rejection based on the reduced-$\chi^2$ of their respective SED fits.}
    \label{fig:image_comp_contrast}
\end{figure*}
\section{Analysis}

\begin{figure*} 
    \centering
    \includegraphics[width=\textwidth]{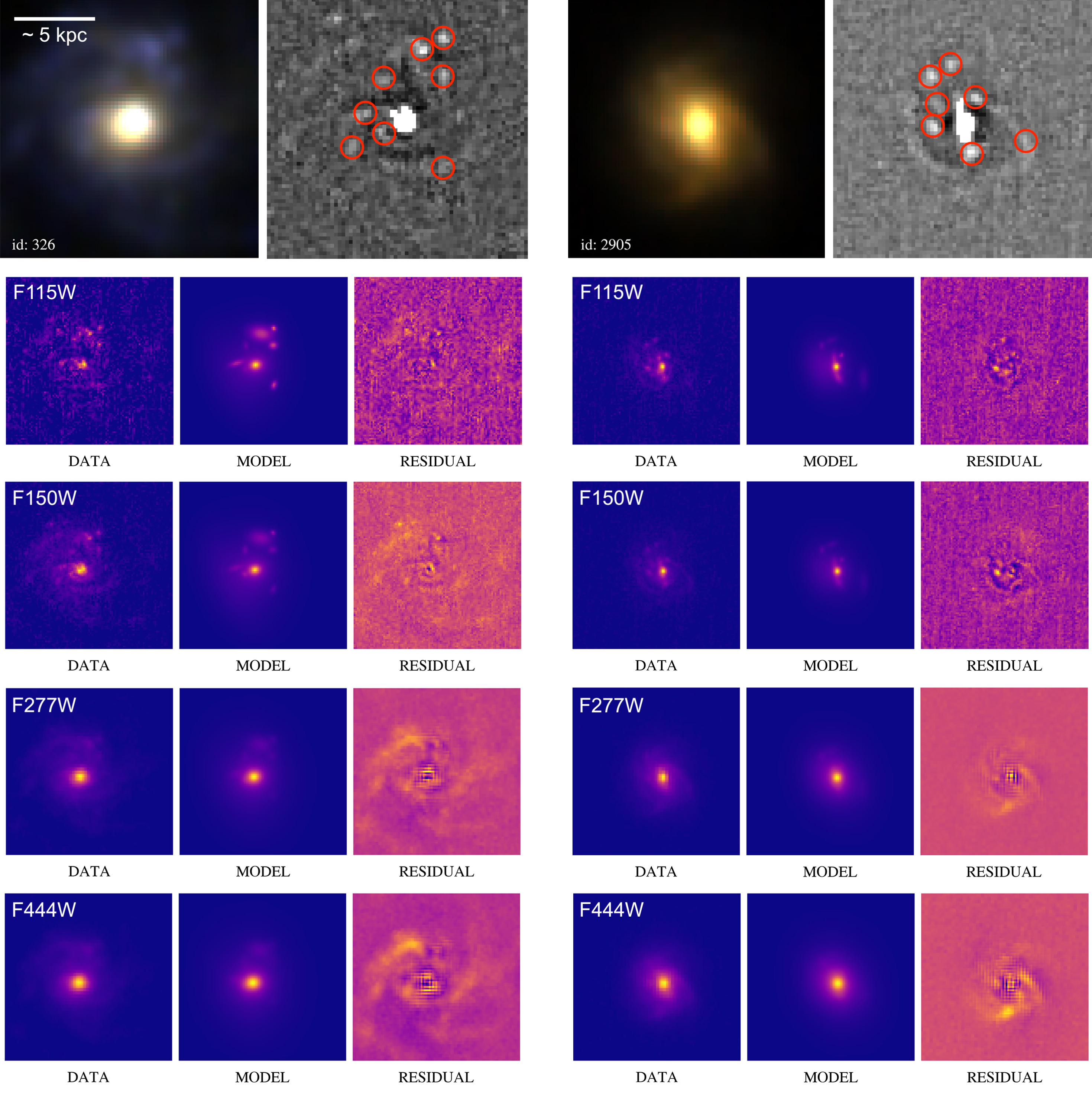}
    \caption{The `deconstruction' of two of the 32 galaxies in our sample. The RGB color image of each galaxy (Fig.~\ref{fig:image_comp}) is shown at the top left. Next to it is the F150W contrast image (from Fig.~\ref{fig:image_comp_contrast}) used to detect the clump locations (initial number of clumps), marked in red. Each clump is associated with a Gaussian model. Below, the corresponding data, model, and normalized residual (data-model/noise) for each JWST/NIRCam image are provided. Note that the color image is created by PSF-matching the F150W (blue) and F277W (green) images to the F444W (red) image, while the deconstruction is done at the native resolution of the respective filters.}
    \label{fig:image_decomp}
\end{figure*}

\begin{figure*} 
    \centering
    \includegraphics[width=0.9\textwidth]{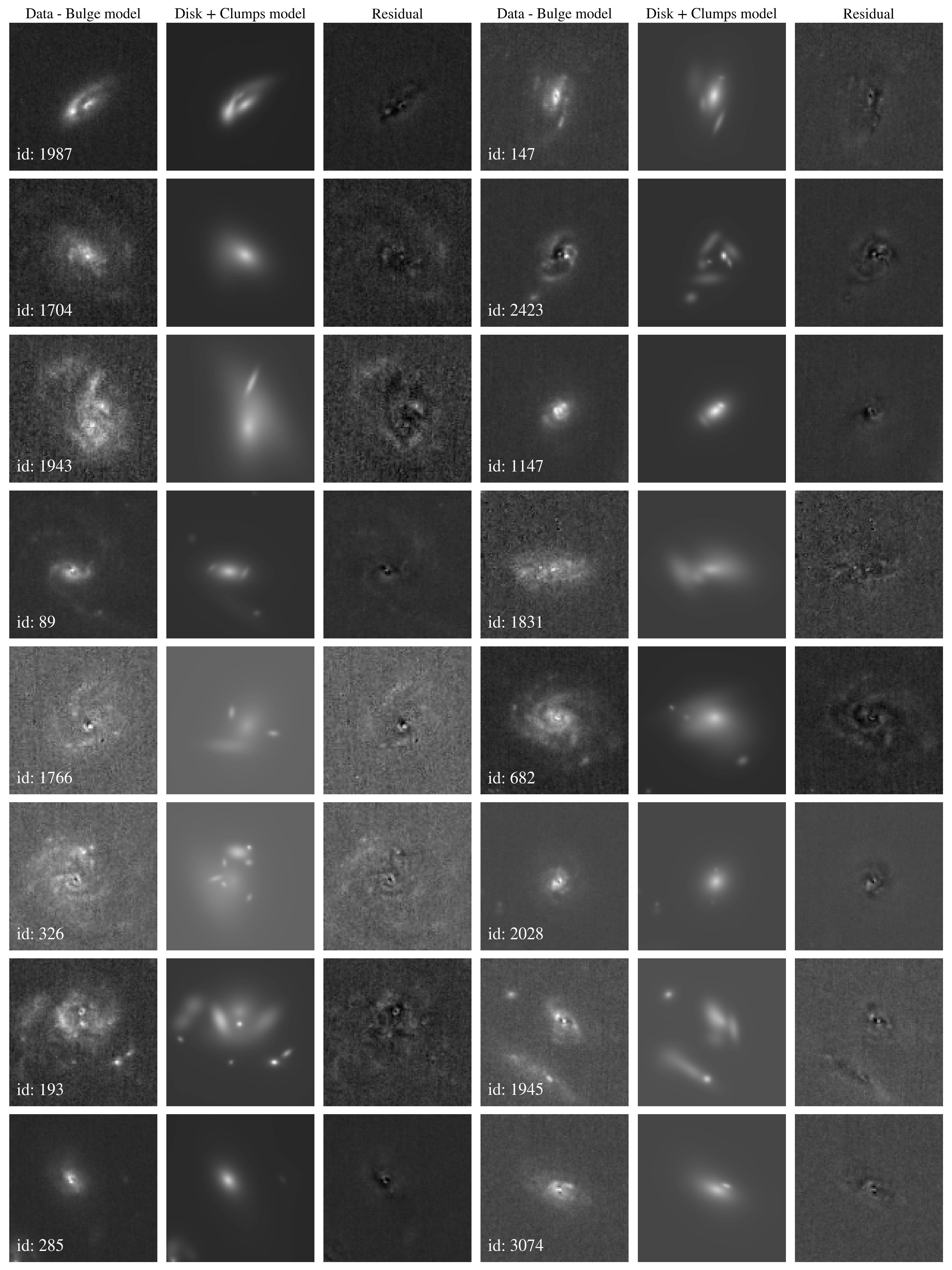}
    \caption{(Data-Bulge model) The F150W image of the clumpy disks,  after the subtraction of the bulge model for 16/32 galaxies in our sample.  (Disk+clumps model) The corresponding model image,  that includes the disk and all initially found clumps,  without the bulge.  (Residual) The final residual image after the subtraction of all the model components from the data.}
    \label{fig:image_decomp_models_1}
\end{figure*}
\begin{figure*} 
    \centering
    \includegraphics[width=0.9\textwidth]{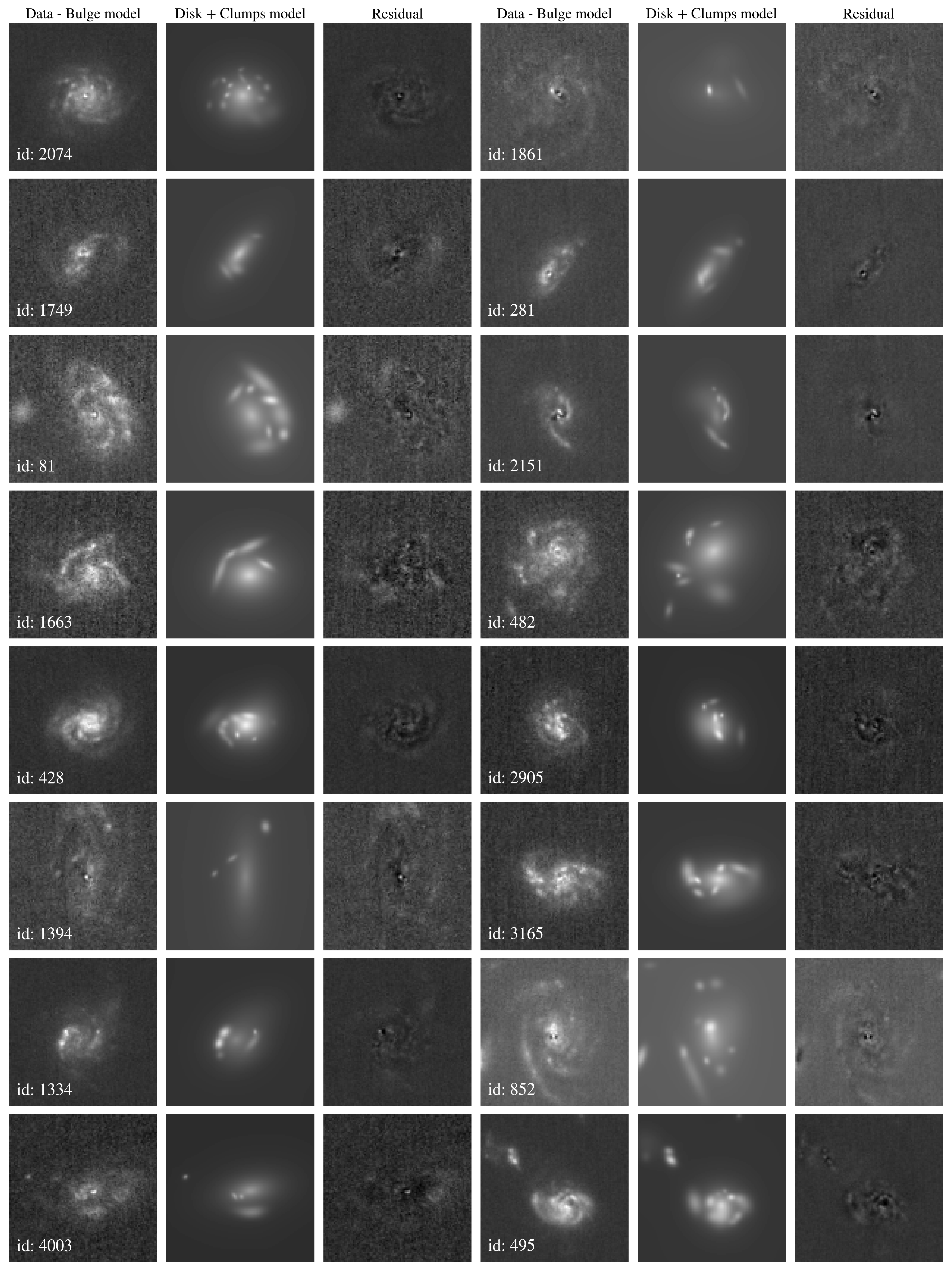}
    \caption{The same images as in Fig.~\ref{fig:image_decomp_models_1} for the remaining 16/32 galaxies.}
    \label{fig:image_decomp_models_2}
\end{figure*}

\subsection{Sample and data} \label{subsec:sample}
For this work, we begin with the FMOS-COSMOS-ALMA sample of 57 star-forming main-sequence galaxies\footnote{AGN hosts are exclused in the sample selection based on broad-line data and X-ray flux} (within a $\sim 0.5\, $ dex scatter; Kashino et al. in prep., Suzuki et al. in prep.), with both JWST/NIRCam and ALMA continuum data. This is a subset of the spectroscopically confirmed sample of over 1500 galaxies with H$\alpha$ detection in the COSMOS field \citep{scoville07b,silverman15,kashino19}. The galaxies are within a redshift range of $1.43 \leq z \leq 1.74$ and have stellar masses of $10^{10.5-11.4}\,\rm M_{\odot}$. Each of these has H$\alpha$ and [NII] emission line detections, allowing measurements of star formation rates (SFRs),  average Balmer decrements (using additonal H$\beta$ detections and upper-limts),  and metallicities \citep{kashino13,kashino19}. Thus, we have access to metallicity-dependent dust-to-gas ratios (Kashino et al. in prep), which will be used in this work. The parent FMOS sample was selected using H$\alpha$, which introduces a potential bias toward low dust attenuation \citep{silverman15}. However, this works in our favor as it allows for better detectability of clumpy structures using optical bands, which is the focus of this study.

Most of these galaxies are covered by the COSMOS-Web survey \citep{casey23}.  11 galaxies do not have coverage in all four bands due to pointing variations,  which are removed from our sample.  This provides us with 4-band JWST/NIRCam data (with average $5\sigma$ AB magnitude depths; PSF FWHM): F115W (27.2 mag; $0.04^{\prime\prime}$), F150W (27.4 mag; $0.05^{\prime\prime}$), F277W (28.0 mag; $0.09^{\prime\prime}$), and F444W (27.9 mag; $0.14^{\prime\prime}$). Each image has a pixel scale of $0.03^{\prime\prime}$. The PSFs for each filter are created using the software \textsc{PSFEx} \citep{bertin11} on the full COSMOS-Web mosaic. We also have access to the F814W HST/ACS imaging \citep{koekemoer07} with a $5\sigma$ depth of 27.2 mag and PSF FWHM of $0.08^{\prime\prime}$. However, we find that the fitting procedure for clumps, as described in Sec.~\ref{subsec:jwst_measurements}, becomes highly uncertain in F814W, which covers the rest-frame UV for our sample. This is due to a combination of low intrinsic flux ($F_{\nu}$) that is not compensated by the already shallow data, along with the broader PSF compared to the adjacent band (F115W). In the fraction ($\sim 50\%$) where we do have reliable clump flux estimates, we confirm that our results do not change with or without using this filter. Hence, we do not include it in the final analysis.

We discard any galaxies with obvious signs of (major-)merging activity by visually inspecting each of the four NIRCam images\footnote{However, the sample galaxies might be experiencing minor-merging activity, which cannot be identified visually.}. This is to ensure that we are primarily targeting clumps arising from dynamical instability in rotationally supported disks. While this is not a particularly aggressive method of assessment, we find that each of the remaining galaxies is well-fit with a single bulge+disk model in the rest-frame near-IR (further discussed in Sec.~\ref{subsec:jwst_measurements}), supporting our assessment that the galaxies are not undergoing highly disruptive major mergers. Using the same bulge+disk modeling, we also reject galaxies with disk axes ratios $<0.5$. This ensures that we study only galaxies where clump detection is not missed due to orientation, restricting our sample to nearly face-on objects.  We are left with 32 galaxies, whose RGB images are shown in Fig.~\ref{fig:image_comp}.  The corresponding contrast F150W images, used for clump detection and shape estimation are provided in Fig.~\ref{fig:image_comp_contrast} (see Sec.~\ref{subsec:jwst_measurements} for details).

Finally, we also use the Atacama Large (sub-)Millimeter Array (ALMA) $870\,\mu$m band-7 continuum data to characterise the rest-frame sub-mm galaxy flux. The $1\sigma$ depth is found to be $\approx 200\,\mu$Jy/beam and the synthesized beam size is $0.69^{\prime\prime} \times 0.57^{\prime\prime}$ if naturally weighted.  These data were taken as part of ALMA proposal ID: 2021.1.01133.S (PI: D. Kashino). It should be noted that here we use UV-space fitting, following methods from previous works \citep[e.g.,][]{puglisi19,kalita22,tan24,tan24b}. This allows us to fit profiles that may have smaller sizes than the synthesized beam, provided the S/N ratio is sufficiently high.

\begin{figure} 
    \centering
    \includegraphics[width=0.48\textwidth]{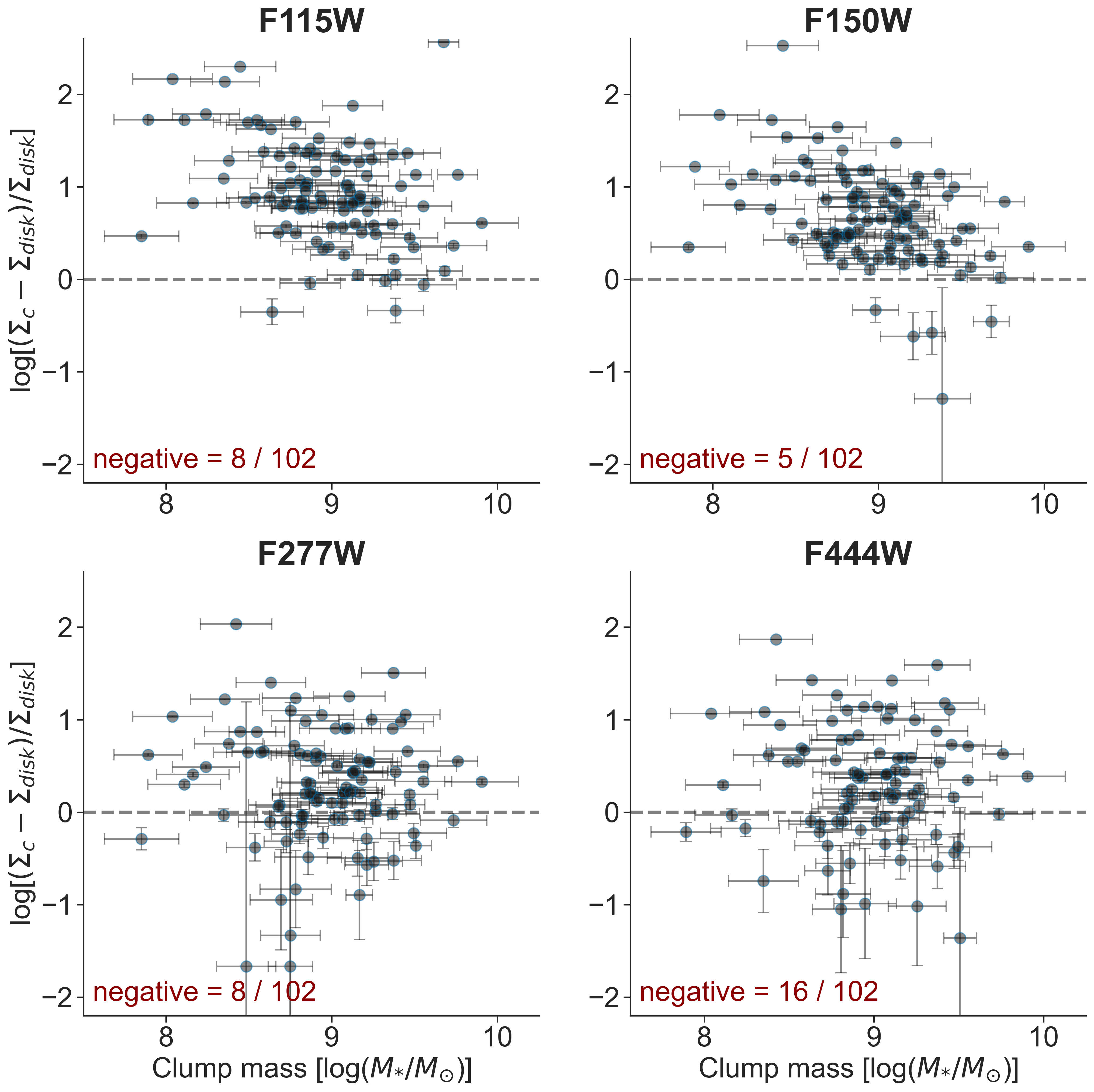}
    \caption{The surface brightness contrast between the clumps (detected in F150W) and the rest of the host disk in different filters as a function of their stellar mass. In some cases (termed `negative'), the surface brightness is lower than the disk's average, which is possible since their detection depends on the local flux distribution around the clump region. Notably, the number of such clumps is higher in the rest-frame near-IR (F277W and F444W) compared to the rest-frame optical (F115W and F150W).}
    \label{fig:clump_detectability}
\end{figure}

\subsection{JWST data measurements: Bulge+Disk+Clumps} \label{subsec:jwst_measurements}

For the analysis of the images, we introduce a new method for fitting the sources, thereby `deconstructing' the galaxies into their constituent bulges, disks, and clumps (Fig.~\ref{fig:image_decomp},\ref{fig:image_decomp_models_1},\ref{fig:image_decomp_models_2}). Based on the analysis of clumpy galaxies in \cite{kalita24, kalita24}, we find that the contrast between the clump flux and the underlying disk flux increases from the rest-frame near-IR (F277W, F444W) to the rest-frame optical (F115W, F150W). This is reflected in the flux surface density contrast relative to the underlying disk (Fig.~\ref{fig:clump_detectability}). This effect can be attributed to two factors: (I) the near-IR traces older stellar populations, thereby mapping the average stellar mass distribution rather than star formation (which is higher in clumps), and (II) the resolution of these longer wavelength bands being a factor of $\gtrsim 3$ lower than that of the shorter wavelength ones.  For the same reasons, the rest-frame near-IR is effective at mapping the underlying bulge and disk \citep{sheth10}.

We employ a two-stage spatial model-fitting procedure that optimizes the bulge and disk models using the F444W filter and the clump models using the F150W filter. We choose to use F150W for clump detection instead of F115W due to the latter's shallower depth and possible biases from an under-sampled PSF.  We start by fitting a dual-Sérsic profile with indices $2$ and $1$ for the (pseudo-)\footnote{We ensure that using a classical bulge with a Sérsic index $=4$ does not change our results. However, we use the value for a pseudo-bulge since the central region is star-forming throughout our sample. } bulge and disk, respectively, in the F444W filter.

Following this, clump modeling is performed by first detecting these structures in the contrast images which are shown in Fig.~\ref{fig:image_comp_contrast} \citep[following procedures presented in][]{kalita24, kalita24b}.  These are created by smoothing the image with a Gaussian kernel with $\sigma = 3\,$pixels,  and then subtracting it from itself. The detection is done in stages using incrementally lower $\sigma$-thresholds to find regions of progressively fainter flux peaks.  At each stage,  the detected clumps are fit with elliptical Gaussian models, along with the existing bulge+disk model from the first stage. As clumps are progressively added, increasing the model's complexity, we simultaneously measure the Bayesian Inference Criterion (BIC) at each step. This allows us to evaluate the trade-off between the goodness of fit and the model's complexity.  For each galaxy,  the lowest BIC values are reached at either $3$ or $4\,\sigma$ threshold. 

The final bulge+disk+clump spatial models (Fig.~\ref{fig:image_decomp_models_1},\ref{fig:image_decomp_models_2}) are fixed (i.e., their respective shapes) and are used, together with their respective PSFs, for deblending all four JWST/NIRCam filters for the final flux measurements\footnote{In \cite{kalita24b},  a background was subtracted, which was estimated using the underlying disk flux at the relevant radial distance from the center while excluding the clumps.  However, this is not necessary here since we are simultaneously modeling the clumps and the underlying disk}. Further details of the procedure, along with caveats, are provided in Appendix~\ref{append:jwst_deconstruction}.  Finally, uncertainties for all flux values are calculated by artificially adding sources of the same size but varied flux values as part of composite models and then remeasuring them. This process is conducted separately for each filter. We then repeat the procedure for size measurements, fixing the flux and varying the sizes. Given that clump size measurements are performed only in the F150W filter, other bands are excluded from this part of the analysis. The levels of uncertainty for F150W flux and sizes can be found in Appendix Fig.~\ref{fig:size_and_light}.

The photometry is passed on to the SED fitting process to derive physical parameters (described briefly in the next section). Based on the BIC values of the fits at different stages, we conclude the following: 
\begin{itemize}[leftmargin=*, noitemsep, topsep=0pt] 
\item Rest near-IR: All 32 galaxies are better fit (BIC-based, accounting for model complexity) with a bulge+disk model in the rest-frame near-IR (F444W) compared to a single Sersic fit. However, adding clumps to the model results in a significant improvement in the BIC ($>10^{3}$) for all galaxies. 
\item Rest optical: The bulge+disk model derived from the rest-frame near-IR does not fit the galaxies well in the rest-frame optical (F115W and F150W) due to the dominant presence of clumps in these images. However, adding the clumps to the model results in a substantial improvement in the BIC values ($>10^{5}$).  The improvement can be attributed to the fraction of the galaxy optical flux contained in the clumps ($\sim 5-30\%$). Thus, the bulge+disk+clump models significantly outperform the bulge+disk models in the rest-frame optical. 
\item All bands: The final comparison of the reduced-$\chi^2$ values across all four NIRCam images shows the trend: $\chi^2_{\rm F115W}$ < $\chi^2_{\rm F150W}$ < $\chi^2_{\rm F277W}$ < $\chi^2_{\rm F444W}$, with lower reduced $\chi^2$ values indicating a better fit. We do not use the BIC values for this comparison since the same model is used for all bands. The increasing reduced $\chi^2$ with wavelength is likely related to residuals in the near-IR, where we clearly detect underlying stellar substructures that resemble spiral arms in most cases. We discuss their relevance later in Sec.~\ref{subsec:spirals_results} and \ref{subsec:spirals_discussions}. Finally,  the shallower depth of the F115W image in comparison to that of F150W,  along with the higher relative the contrast between the clumps and the disk (Fig.~\ref{fig:clump_detectability}),  and the higher resolution likely resolving out extended residual substructures, results in lower reduced $\chi^2$ in F115W. 
\end{itemize}

\subsection{ALMA data measurements: Bulge+Disk} \label{subsec:alma_measurements}

The second part of the analysis uses the same principle of galaxy image deconstruction on the ALMA $870\,\mu$m image. Given the synthesized beam size of $\sim 0.6^{\prime\prime}$, we find that the signal-to-noise ratio of the clumps is insufficient for accurate measurements. Therefore, we do not measure the $870\,\mu$m flux for individual clumps. However, it is sufficient for robust flux measurements of the bulge and disk through deblending (details are provided in Appendix~\ref{ap:alma_meas}).  Since we do not deconvolve the clumps here, the disk flux includes the clump contributions.

We use the bulge+disk model derived from the F444W rest-frame near-IR image, with the position and shape fixed, to extract the $870\,\mu$m flux. This is done by translating the image plane parameters to the UV-plane \citep[using values from][]{tan24} and then measuring the flux of each (fixed) component. We use this approach because the ALMA data does not have sufficient signal-to-noise ($>50\,\sigma$) for robust shape measurements \citep{tan24}. Here, we assume that the stellar light and dust emission are coincident, which has been supported by recent studies \citep{liu24, hodge24}. \cite{tan24b} also concludes that part of the sub-mm flux can be attributed to the star-forming, under-construction bulge. Using the bulge+disk model allows us to determine the bulge and disk contributions to the total $870\,\mu$m flux from the galaxy.

We compile the bulge and disk photometry using the four JWST/NIRCam bands and ALMA Band-7. The disk flux in the NIRCam bands includes the flux of all clumps within the $90\%$ flux radius ($r_{90}$) of the disk; any flux beyond this radius is not considered.  A similar approach is unnecessary for ALMA, as we deblended the data using only a bulge+disk model that includes any clump flux. These flux measurements are then used in an SED analysis (Appendix~\ref{append:sed_fitting}) to calculate physical properties such as stellar mass and SFRs.  The analysis uses constant star-formation history models to fit the 5-band photometry for the bulge and disk of each galaxy using the Python-based ``Code Investigating GALaxy Emission" \citep[CIGALE;][]{boquien19} tool. The clump SED fitting relies only on the four JWST bands, as their ALMA band-7 flux is unavailable. The $1\sigma$ errors in the measured parameters are based on the $\Delta \chi^{2} = \pm1$ of their respective distributions as reported by CIGALE.

We already have the integrated dust mass of the galaxies from the 44-band \citep[rest-frame UV to far-IR COSMOS data;][]{weaver22} SED fitting, which is converted to gas mass using metallicity-dependent dust-to-gas ratios (to be presented in Kashino+in prep). Assuming similar SEDs for the bulge and disk, we use the ratio of the $870\,\mu$m fluxes to estimate the gas mass of the bulge and disk. These values will be used later in Sec.~\ref{subsec:galaxy_prop}.

\subsection{Detection completeness and uncertainties} \label{subsec:completeness}

\begin{figure} 
    \centering
    \includegraphics[width=0.48\textwidth]{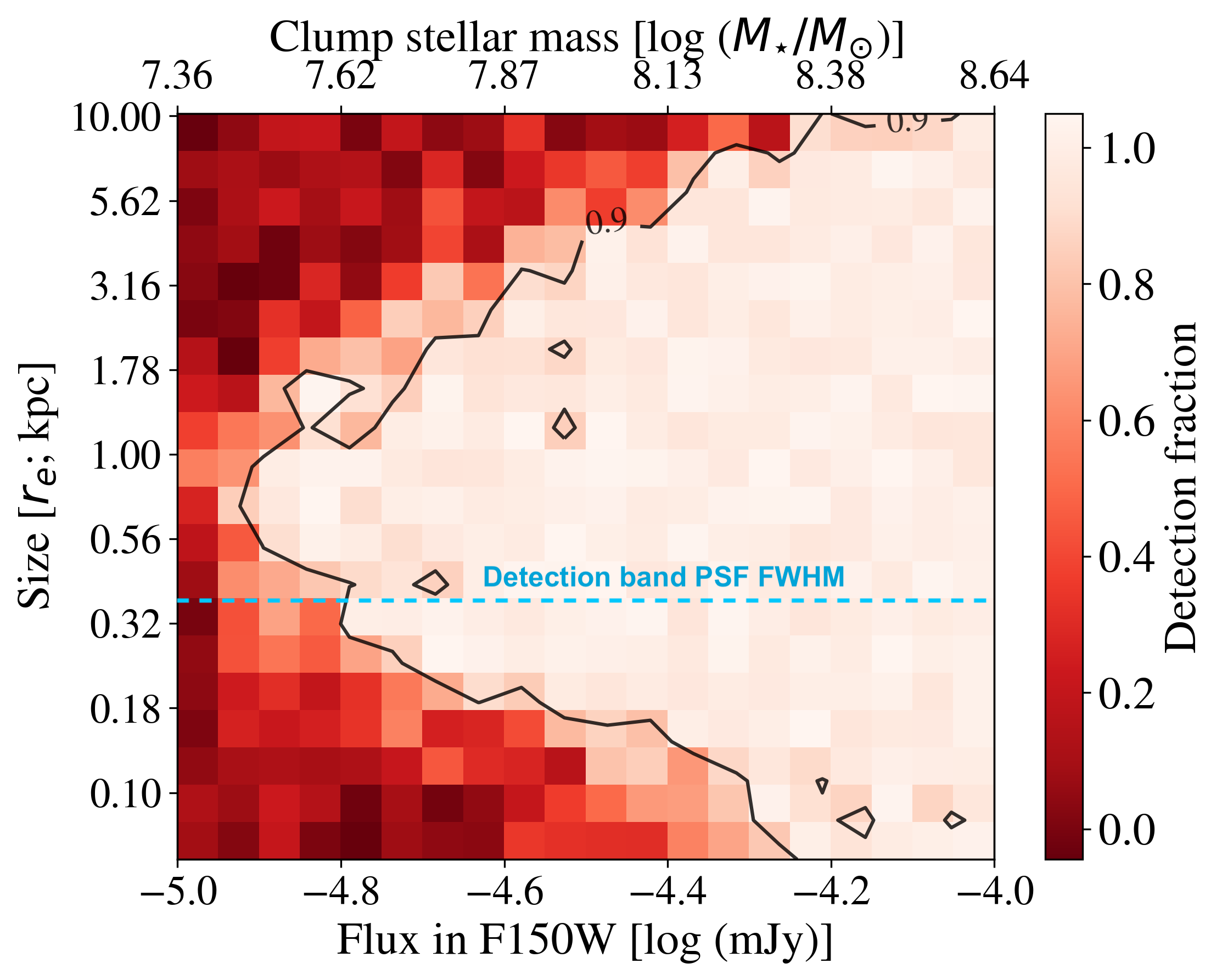}
    \caption{The completeness of our detection algorithm as a function of clump F150W flux (detection filter) is shown on the bottom x-axis, and clump size ($r_{e}$). The corresponding clump mass is computed using the average mass-to-light ratio of our clump sample and displayed on the right as a function of size on the top x-axis. The physical size of the PSF (FWHM) of the detection image (F150W) is also indicated in blue.}
    \label{fig:completeness}
\end{figure}

The first test of the robustness of the analysis is whether we account for the total stellar mass and SFR of the galaxies. We therefore compare the total stellar mass and SFR of the bulge and disk of each galaxy (the latter also including the clumps) to the integrated values for the galaxies presented in Kashino et al. (in prep.). They use a similar SED fitting setup as in our work (mentioned in Sec.~\ref{subsec:alma_measurements} and details in Appendix~\ref{append:sed_fitting}) but have additional access to the full wavelength coverage of the COSMOS field \citep{weaver22}. We find agreement for the SFR values within the errors, while there is, on average, a $0.1\,\rm dex$ offset in stellar mass, with our values being lower (Appendix Sec.~\ref{append:sed_fitting} and Fig.~\ref{fig:kashino_comp}). We attribute this to the use of $3^{\prime\prime}$ apertures in Kashino et al. (in prep.), which are characteristically larger than the extent of our galaxy model (the measured disk $r_{90}$ is on average $\sim 1.5^{\prime\prime}$). However, it should be noted that we miss some flux within the galaxies in the longest bands, which appears in the form of spiral arms in the residuals. Exactly estimating the flux in these residuals is challenging since they contain both positive and negative flux values due to over-subtraction and under-subtraction at different locations.

The detectability of the clumps in the galaxies would likely depend on their intrinsic flux and size, which contribute to the contrast with the underlying disk. We therefore artificially add clumps to the galaxies in the corresponding F150W image, spanning a reasonable range of fluxes and sizes, and attempt to detect them using the same procedure as discussed in Sec.~\ref{subsec:jwst_measurements}.  We randomly place these clumps over the surface of the disk (up to the $r_{90}$) while ensuring they do not fall within the bulge region\footnote{It is challenging to determine the exact extent of the bulge to ensure the artificial clumps are not placed within it. We use the segmentation map in F150W, which is used to detect the real clumps. Here, the bulge is also detected as a clump, which is usually rejected in our analysis. However, the associated segmentation region provides an estimate of the bulge's extent.}.  

We consider a clump as detected if the procedure finds a new source within the effective radius of the artificial clump. This gives us the detection completeness of our procedure as a function of F150W flux and size (Fig.~\ref{fig:completeness},  bottom x-axis).  We carry out this procedure for each galaxy, with the final result representing the average across the full sample. It is worth noting that we do not re-fit the artificially added clumps if they are detected. This decision is made because the accuracy of the fitting procedure in extracting correct flux and size measurements depends not only on clump properties but also on variations in the underlying disk flux. Thus, error quantification is performed for each clump and has been discussed in Sec.~\ref{subsec:jwst_measurements}. The goal of determining completeness here is simply to gain a general understanding of clump detectability in our study.

Since we already measure the stellar mass of the clumps in our sample, we estimate a mass-to-light ratio\footnote{$\rm \log(M_{\star}) = (\log (flux) + 10.74\pm1.94)/(0.78\pm0.21)$} from the corresponding F150W flux measurements (relation shown in the Appendix, Fig.~\ref{fig:mass_to_light}).  We then use this to estimate completeness in terms of size and stellar mass (Fig.~\ref{fig:completeness},  top x-axis).  It is worth noting that the flux-to-stellar mass conversion is accurate only up to the uncertainty of the SED-based stellar mass estimation, which is on average $\sim 0.35\,$dex.  However,  the uncertainty of the representative conversion for the full clump sample would also depend on the deviation of the mass estimates from the average ratio (Fig.~\ref{fig:mass_to_light}).  Hence we settle on a conservative uncertainty of $0.5\,\rm dex$.  For the smallest clumps ($\sim 0.1\,\rm kpc$) with a completeness flux limit of log(Flux$_{\rm F150W\,|\,mJy}$) $= -4.3 $ we therefore find a resulting stellar mass limit of $\approx 10^{8.7}\,\rm M_{\odot}$.  The additional $0.5\,\rm dex$ improves our completeness estimation by addressing the previously unaccounted robustness of the flux extraction. These uncertainties are also similar to the systematic variation we might expect in clump mass measurements as a function of galactocentric radius \citep[$\sim 0.35\,$dex that is estimated in ][]{kalita24b}. Hence, clump detection completeness has not been characterized as a function of radial distance from the center in this study.

The resolution of the F150W image (FWHM $\sim 0.4\,\rm kpc$ at $z \sim 1.5$) influences the detection completeness of the clumps. However, we detect and measure clump sizes down to $\sim 0.1\,\rm kpc$ due to the high signal-to-noise of these structures. Given that we are fitting models, we can therefore characterize sub-PSF scales (as mentioned in Sec.~\ref{sec:jwst_clumps}), albeit with relatively higher uncertainties. This effect can be observed in Fig.~\ref{fig:completeness},  where the smallest clumps can be detected only when their flux (and hence signal-to-noise) is high. Nevertheless, throughout the work, we ensure that our results are valid regardless of whether we include or exclude the sub-PSF clumps.

Another key feature of Fig.~\ref{fig:completeness} is the decrease in detection fraction for the largest sizes. This is primarily due to the flux being spread over a larger surface area, which reduces the signal-to-noise ratio of each pixel, making detection more challenging. Additionally, this characteristic, observed above a few kpc, is due to the clump sizes approaching those of the galaxy disks, which have a $90\%$ flux radius of $\sim 5-10\,\rm kpc$. Such clumps would not be identified as localized flux increments necessary for detection. Meanwhile, the smallest clumps may go undetected if they have low flux, as they are excluded by the detection algorithm when detected over $<5$ pixels. This criterion is implemented to reject noise peaks.  The decreasing completeness limit below the PSF size of $\sim 0.4\,\rm kpc$,  as evident from Fig.~\ref{fig:completeness}, is a result of this rejection.

Finally, one caveat of our procedure that must be noted is the complex structure of clumps, which we attempt to fit using a Gaussian model. The goal is to capture the most dominant structure in terms of net flux, which would be represented in the final best-fit model. However, in multiple cases, we observe that the residuals still contain smaller substructures not accounted for by the clump models. Ideally, we would add additional models at the locations of these substructures, in addition to the primary clump model. However, this approach often results in a failure to converge or no improvement in the final BIC. Therefore, we acknowledge this as a limitation that cannot be resolved within the scope of our data and procedure.

\section{Results}

\begin{figure} 
    \centering
    \includegraphics[width=0.48\textwidth]{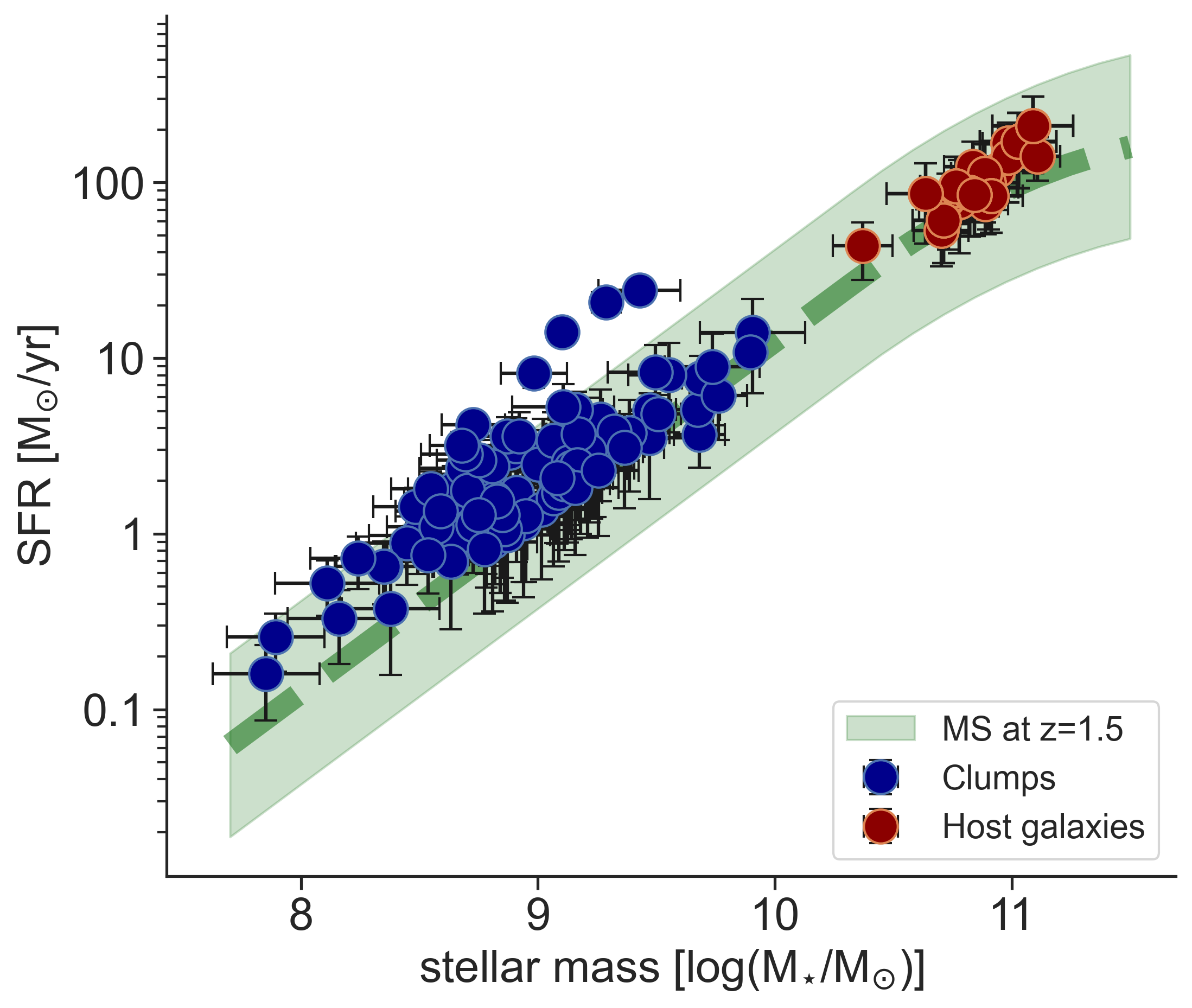}
    \caption{The SFR vs. stellar mass relation of the clumps in our sample, along with that of the host galaxies, is shown.  For reference, the star-forming main sequence at $z=1.5$ \citep{schreiber15}, with a $\pm 0.3~$dex scatter region, is also provided. All clumps are found to lie within the scatter of this relation, though generally above the average. Meanwhile, the hosts, originally selected to be within the main sequence, unsurprisingly follow the relation more closely.}
    \label{fig:clump_ms}
\end{figure}

\begin{figure} 
    \centering
    \includegraphics[width=0.45\textwidth]{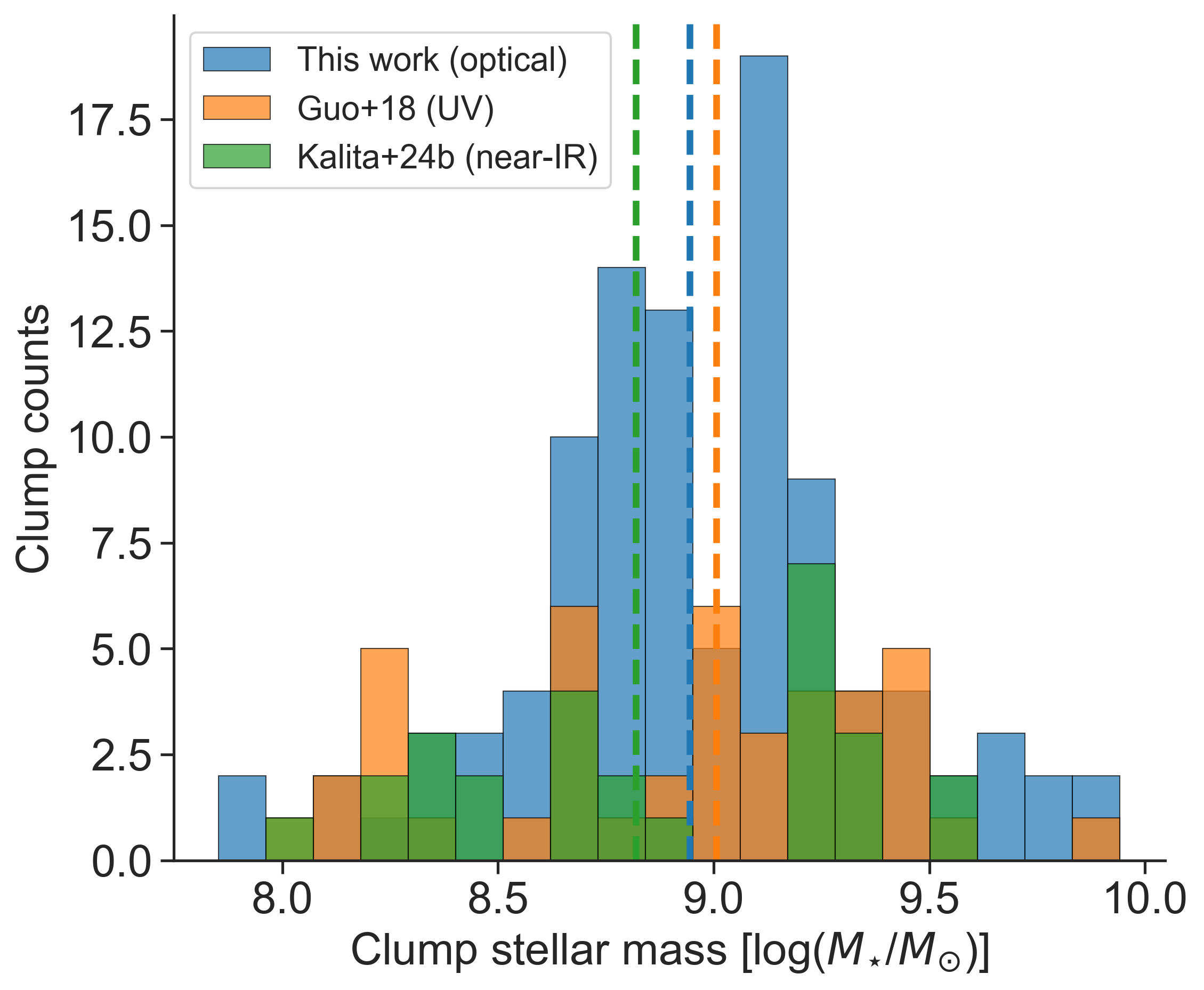}
    \caption{The distribution of stellar mass of the clumps in galaxies falls within the stellar mass range of our sample ($ 10^{10.5-11.4}\,\rm M_{\odot}$). For comparison, we also show the same for clumps from two previous studies that provide statistical assessments of clumps in stellar-mass complete samples of galaxies \citep{guo18, kalita24b}. The former selects clumps in the rest-frame UV, while the latter does so in the rest-frame near-IR.  The median values for each dataset are also shown as dashed lines.}
    \label{fig:clump_mass_lit}
\end{figure}

\begin{figure} 
    \centering
    \includegraphics[width=0.49\textwidth]{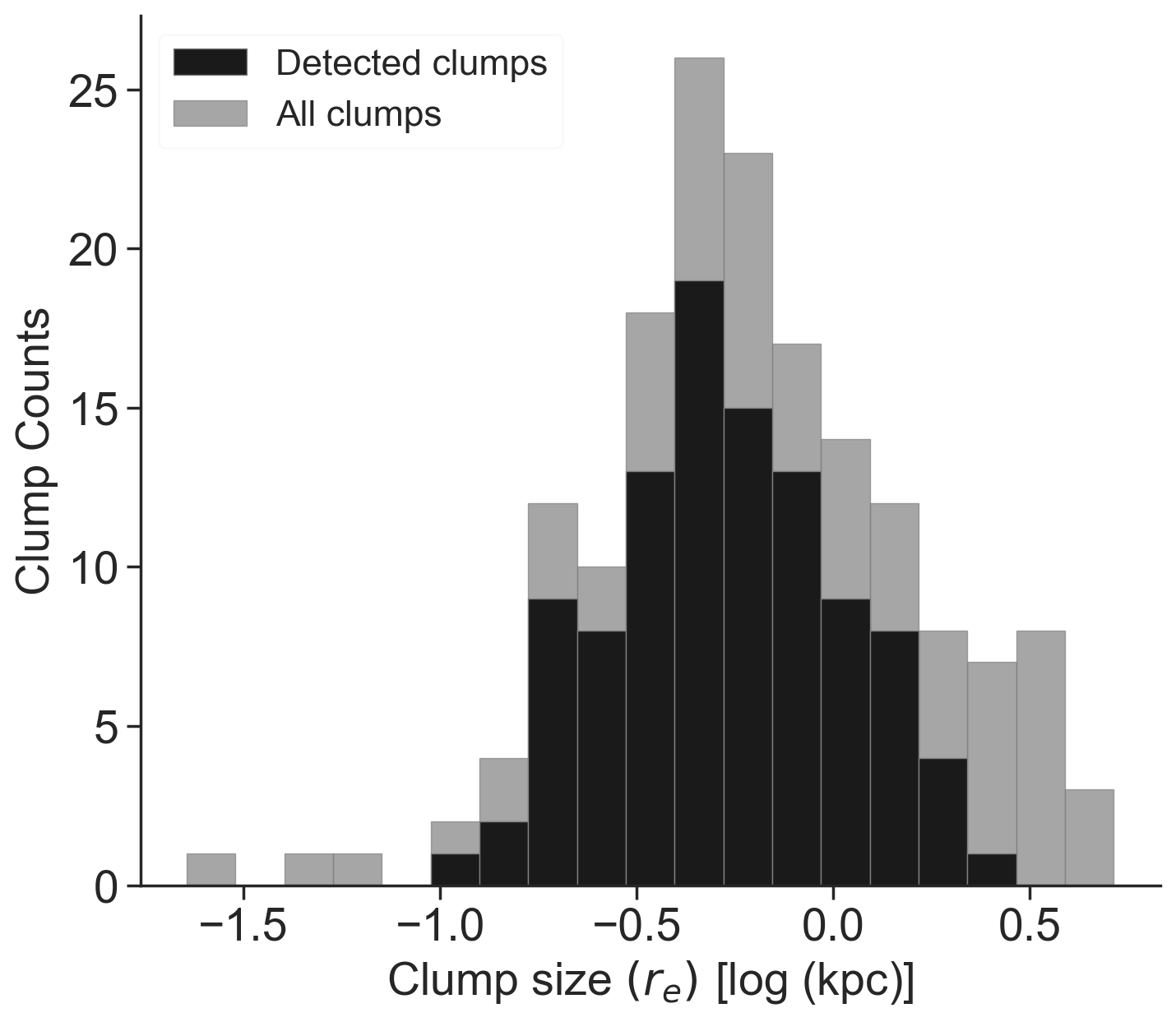}
    \caption{The distribution of clump sizes within our sample.  The `detected clumps' refer to the clumps we include in our study,  while 'all clumps' also include those that have been removed due to a reduced-$\chi^2 > 4$.}
    \label{fig:clump_size_hist}
\end{figure}

\begin{figure*} 
    \centering
    \includegraphics[width=0.99\textwidth]{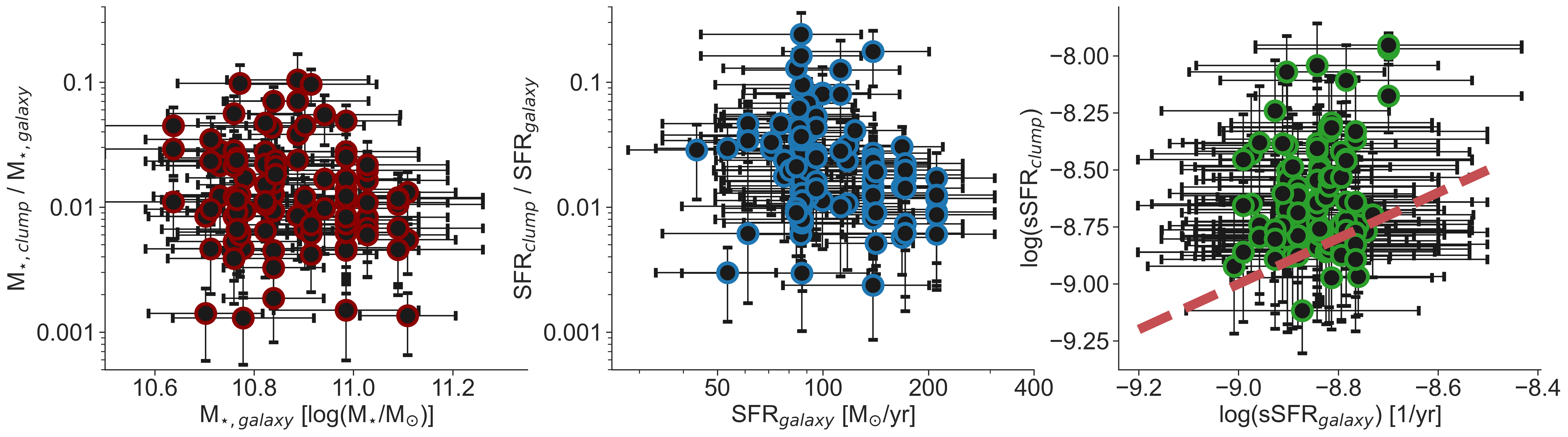}
    \caption{(Left) The ratio of stellar mass within each clump to that of the host galaxy as a function of the host's stellar mass. (Middle) The same ratio, but for SFR. (Right) The sSFR of the clumps vs. that of the host. The red line indicates where clumps would lie if they had the same sSFR as their respective hosts; almost all of our sample is found above this line.}
    \label{fig:clump_ind_ratio}
\end{figure*}

\begin{figure} 
    \centering
    \includegraphics[width=0.4\textwidth]{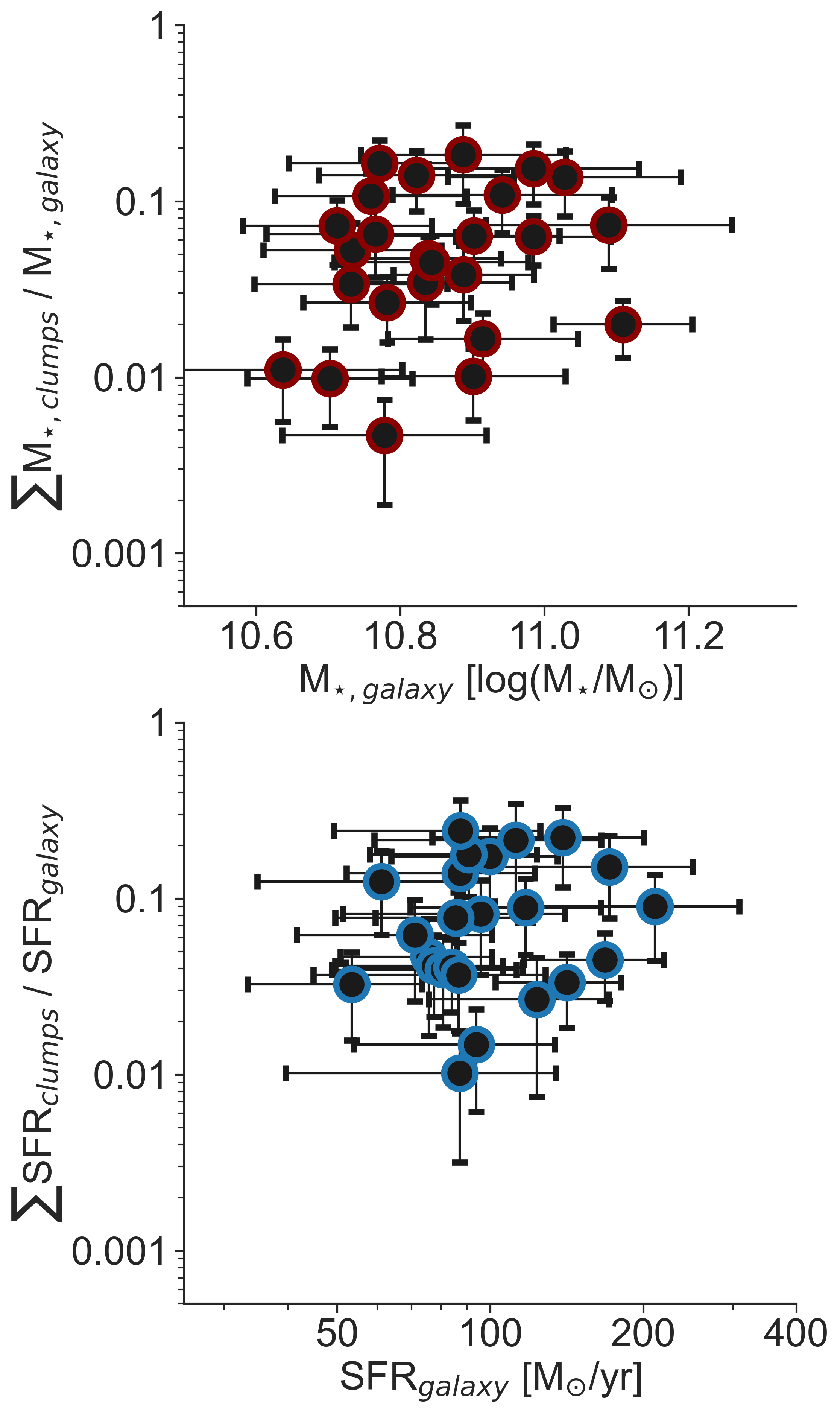}
    \caption{(Top) The ratio of the total stellar mass in the modeled clumps of a galaxy to that of the host galaxy. (Bottom) The same ratio for SFR.}
    \label{fig:clump_tot_ratio}
\end{figure}

\begin{figure} 
    \centering
    \includegraphics[width=0.45\textwidth]{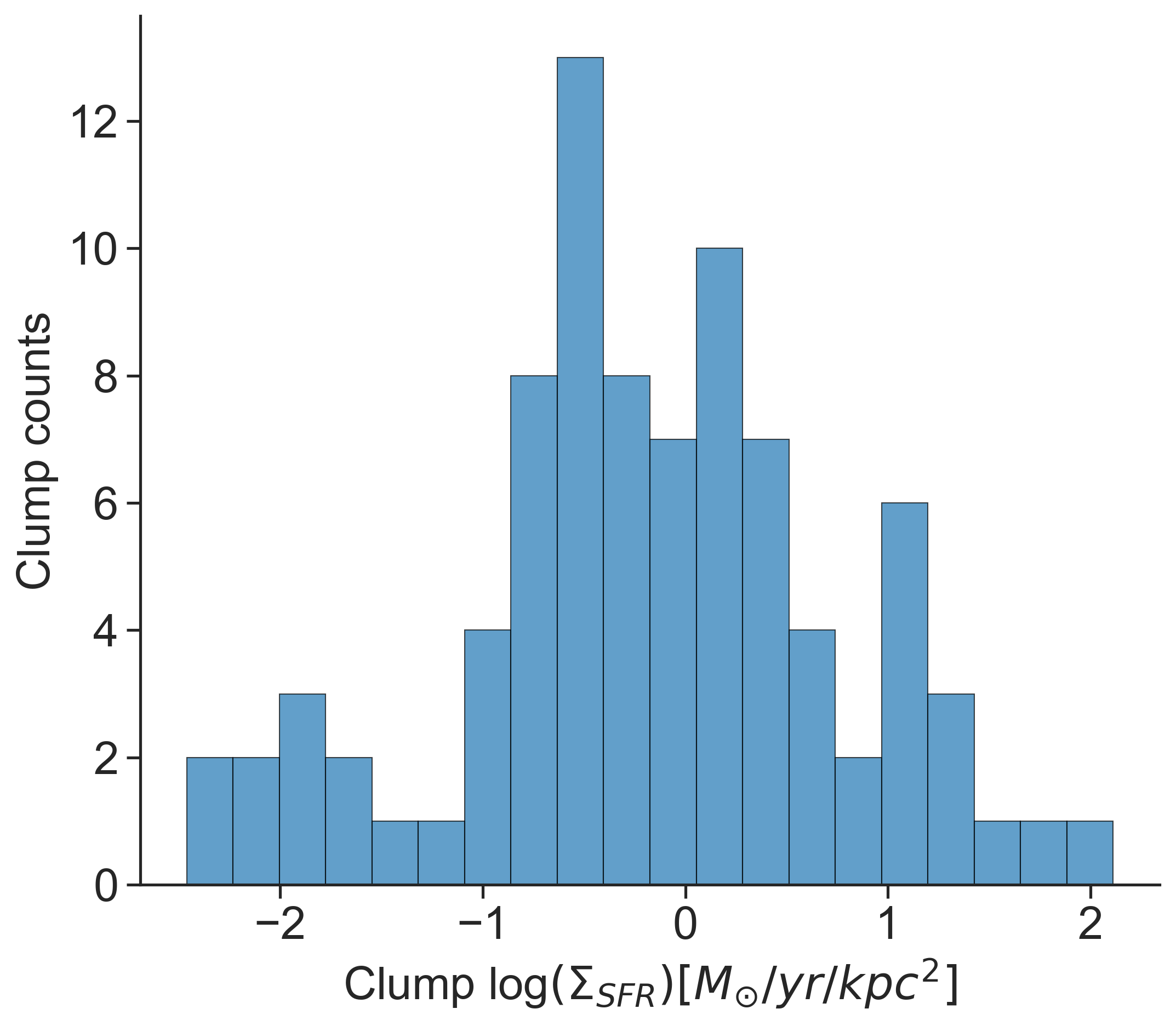}
    \caption{The clump SFR surface density ($\Sigma_{\rm SFR} = \rm SFR/\pi\,r_{90}^2$) distribution for our sample.  }
    \label{fig:clump_sigma_sfr}
\end{figure}

\subsection{Galaxy properties} \label{subsec:galaxy_prop}

Here, we briefly mention a few key results from the bulge and disk model-based measurements. The current study focuses on the clumps within the galaxies. Discussions of global properties, central regions, and bulge-disk properties will be provided in forthcoming works.  We find that the galaxies span a stellar mass range of $10^{10.6-11.2}\,\rm M_{\odot}$ and an SFR of $50-200\,\rm M_{\odot}/yr$ (Fig.~\ref{fig:clump_ms}). As mentioned in Sec.\ref{subsec:jwst_measurements} and Sec.~\ref{ap:bulge_disk},  each galaxy is well-fit by a bulge-disk model (and better than a single Sersic, after accounting for model complexity). This suggests that our sample consists of galaxies with a clear disk and, to varying degrees, a bulge.  Major mergers have thus been excluded (Sec.~\ref{subsec:sample}). The bulge-to-disk stellar mass ratio ranges from $0.1-0.6$.

In each case, the net star formation in the disk is higher than in the bulge, likely fueled by molecular gas contents, characterized by disk gas fractions ($f_{g} = 0.1-0.8$).  These are estimated using the $870\,\mu$m flux based estimations of gas mass discussed in Sec.~\ref{subsec:alma_measurements}.  However, among the 26 galaxies featuring clumps, the minimum $f_{g}$ is $0.36$.  Finally, the disk mass surface density, $\Sigma_{\rm disk} = 10^{8.3-9.3}\,\rm M_{\odot}/kpc^{2}$, accounts for the contribution of both the gas and the stellar mass in the disk\footnote{estimated using the ratio of $0.9\times$ stellar+gas mass and $\pi \times r_{90}^{2}$ }. 

\subsection{Clump properties} \label{subsec:clump_prop}

The model-fitting technique we employ allows us to measure the clump fluxes and sizes\footnote{we use the major-axis value in order for projection effects to not influence our results}, along with their associated uncertainties and completeness limits. Across the sample of 32 galaxies, we detect a total of 167 clumps. In order to exclude interlopers,  this number does not include any sources that might have been detected beyond the rest-frame near-IR $90\%$ flux radius of the disk,  which is taken from the bulge-disk decomposition discussed in Sec.~\ref{subsec:jwst_measurements}.  Only 102 of the clumps have reduced-$\chi^{2}$ values $<4$ of their SED fits. We discard the rest due to the unreliability of their measured properties\footnote{the specific value of reduced-$\chi^{2} = 4$ is chosen as a cut-off since beyond this we find the SED based stellar mass uncertainty to suddenly change from a consistent $\sim 0.35~$dex to $>0.5~$dex.} but still investigate them to ensure none of our key results change if they are included. We will mention them throughout the paper whenever relevant. 

Within this final sample,  the clump stellar masses vary within $10^{8.0-9.5}\,\rm M_{\odot}$ (Fig.~\ref{fig:clump_mass_lit}) with an average uncertainty of $\sim 0.35~$dex ($1\sigma$ uncertainties based on $\Delta-\chi^{2} = \pm 1$).  Meanwhile their half-light radii ($r_{e}$) ranges from $\sim 0.1$ to $\sim 2$ kpc (Fig.~\ref{fig:clump_size_hist}).  We also confirm that the parameter space covered by these clumps is within the $90\%$ completeness limit of our detection procedure (Sec.~\ref{subsec:completeness}; Fig.~\ref{fig:completeness}).  Finally, we find attenuation values $\rm A_{V} \lesssim 0.6\,\rm mag$ and half-mass ages $\lesssim 700\,\rm Myrs$\footnote{using the upper limit of the $68\%$ confidence interval} (the time since the formation of half the stellar mass), reflecting our detection in the rest-frame optical. However, the current data do not provide sufficient accuracy to measure these parameters more precisely and hence do not make any scientific claims based on these values.

The final 102 clumps are distributed among 26 of the 32 galaxies in our sample. Three additional galaxies also have detected clumps, but the SED fitting does not return fits with reduced-$\chi^{2} < 4$.  This high fraction of clumpy galaxies is not consistent with previous studies \citep[$\sim 40\%$ e.g.,][]{guo15,kalita24b},  even after accounting for the difference in methodology\footnote{These works do not use the rest-frame optical bands for clump detection. Nevertheless,  we find that variation of detection bands is far from sufficient to explain the factor of $\sim2$ increase in the fraction of galaxies hosting clumps. Furthermore,  the resolution at which the clumps are detected is lower by a factor of $\sim 2$.  \cite{guo15} also has the additional requirement of a clump needing to have $>8\%$ of the Galaxy UV flux.  Incorporating these,  but relaxing the reduced-$\chi^2$ cut-off we introduce still results in 25/32 galaxies being identified as clumpy.}.  We attribute the higher clumpy fraction to our SFR ($\rm H\alpha$) based selection. Galaxies below the main sequence, heavily dust-obscured and compact \citep[e.g.,][]{elbaz18,gomez-guijarro18}, or bulge-dominated and therefore less clumpy \citep{kalita24}, would not be included in our sample.

We find that the SFR and stellar mass of the clumps are tightly correlated, consistent with the extrapolation of the star-forming main sequence \citep[Fig.~\ref{fig:clump_ms},  with the main sequence from][]{schreiber15}.  Although there is a tendency toward higher SFR values, they are generally within 0.4 dex scatter of the relation. This characteristic is evident in the increased specific SFR (sSFR = SFR/M$_{\star}$) compared to the host galaxy (Fig.~\ref{fig:clump_ind_ratio}, right) where we find $0 \lesssim \Delta \log \rm sSFR \lesssim 0.4$. Meanwhile, individual clumps contribute $0.1-10\%$ of the stellar mass and $0.2-30\%$ of the SFR of their hosts \citep[Fig.~\ref{fig:clump_ind_ratio}, left and middle panels; in agreement with previous works, e.g.,][]{wuyts12, kalita24b}. The total contributions from clumps per galaxy are found to be uniformly distributed over $0.5-20\%$ and $1-30\%$, respectively (Fig.~\ref{fig:clump_tot_ratio}). It should be noted that clumps rejected due to high $\chi^2$ in SED fitting are not included, so the actual contributions could be higher.

We compare the clump stellar masses to \cite{guo18} and \cite{kalita24b}, as these provide the largest samples in recent literature with SED measurements (Fig.~\ref{fig:clump_mass_lit}). Only galaxies within our sample's stellar mass range are included ($ 10^{10.5-11.4}\,\rm M_{\odot}$). While we agree on the limits, we find a relatively large number of clumps with stellar masses around $\sim 10^{9}\,\rm M_{\odot}$. However, we do not make any claims about the mass distribution due to our small sample size of 32 galaxies.  Both studies indicate that higher mass galaxies host higher mass clumps \citep[also suggested by][]{dessauges17,claeyssens24}, potentially reflecting clump mergers that form more massive clumps. This process could lead to a hierarchical distribution of clumps across various masses.

With the measured sizes discussed in Sec.~\ref{subsec:clump_mass_size},  we estimate the SFR surface densities\footnote{calculated using the total SFR of the clumps and dividing by $\pi\,r_{90}^2$. Here $r_{90}$ is the $90\%$ F150W flux radius of each clump, estimated as $1.67 \times r_{e}$, assuming the clumps have a Gaussian profile.}, $\Sigma_{\rm SFR} = 10^{-2}-10^{2}\,\rm M_{\odot}/yr/kpc^{2}$ (Fig.~\ref{fig:clump_sigma_sfr}).  These values are higher than those expected in local galaxies, where $\Sigma_{\rm SFR}$ may vary between $10^{-4} - 10^{-1}\,\rm M_{\odot}/yr/kpc^{2}$ \citep[e.g.,][]{bigiel08,leroy08,rahman12}. However, such values are reached in the integrated $\Sigma_{\rm SFR}$ of starbursts \citep[e.g.,][]{kennicutt21}, although it is unclear if the same applies to individual clumps within these galaxies \citep{hinojosa16}. Meanwhile, lensed galaxy samples and local high-z clumpy galaxy analogues show similar ranges as our sample \citep[e.g.,][]{livermore12,fisher17,messa24b,claeyssens24}. High-resolution ALMA studies of high-z starbursts have also found comparable values of $\Sigma_{\rm SFR}$ \citep{sharda18}.

\begin{figure} 
    \centering
    \includegraphics[width=0.48\textwidth]{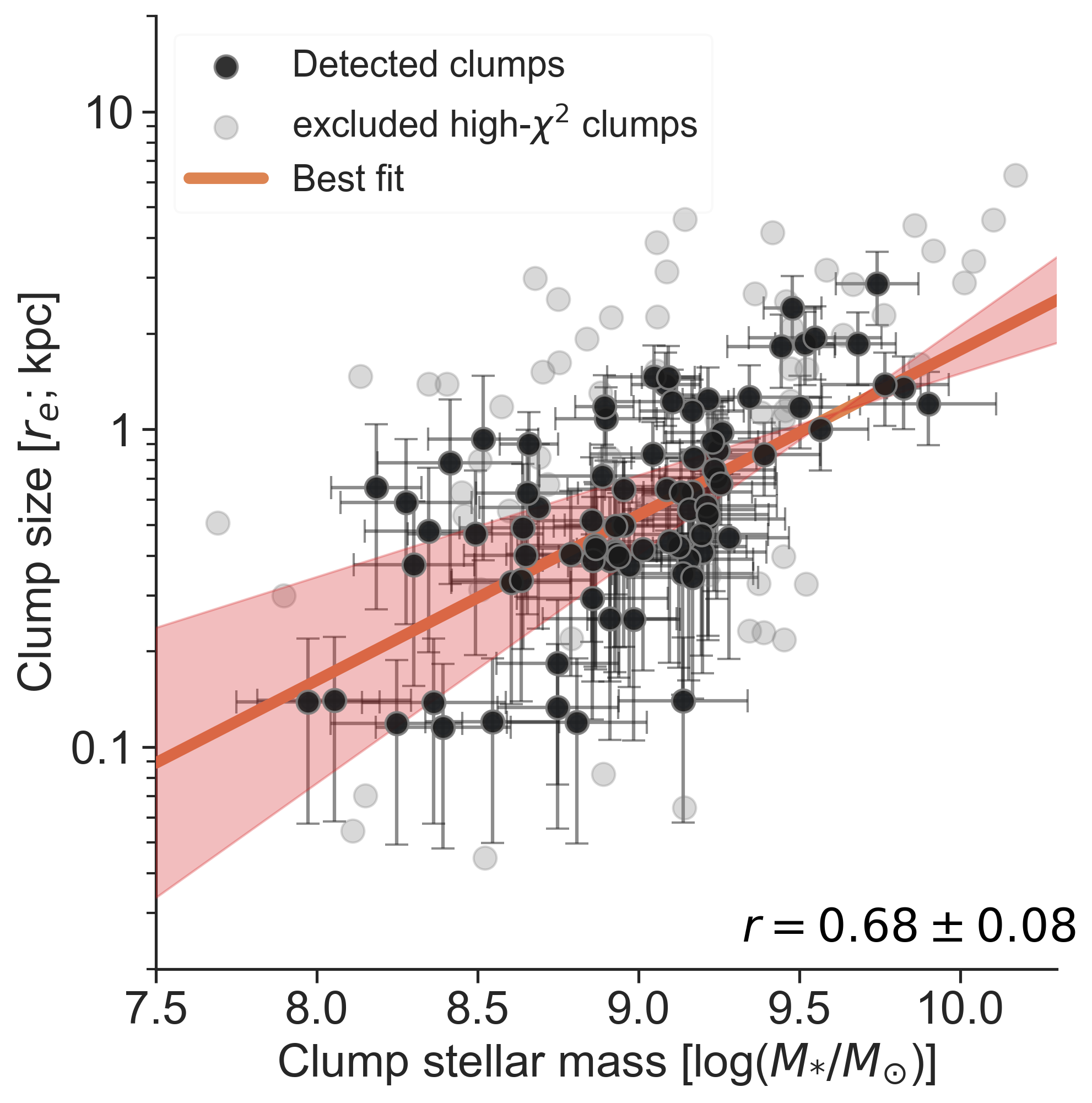}
    \caption{The clump mass-size distribution for our sample, along with the best-fit relationship (in red). The correlation coefficient is $0.68\pm0.08$, with the error estimated through bootstrapping. The high value ($>0.5$) indicates a strong correlation. The grey points represent clumps excluded from our study due to a high reduced-$\chi^{2} > 4$. }
    \label{fig:clump_mass_size}
\end{figure}

\subsection{Clump mass-size relation} \label{subsec:clump_mass_size}

To relate the stellar mass and sizes of the clumps in our sample, we need to ensure that it is estimated within the completeness limits of our procedure.  For clump size $\sim 0.1\,\rm kpc$,  we have already determined a conservative stellar mass limit of $10^{8.7}\,\rm M_{\odot}$ (Sec.~\ref{subsec:completeness}).  However,  for the purposes of deriving a relation,  we exclude clumps with $r_{e}$ below $0.2\,\rm kpc$ that also allows for the inclusion of clumps down to a stellar mass of $10^{8.3}\,\rm M_{\odot}$ (Fig.~\ref{fig:completeness}, with the additional mass estimation uncertainty of $0.5\,\rm dex$ based on Fig.~\ref{fig:mass_to_light}).  Thus,  using the half-light radius and stellar mass of the clumps within this parameter space,  we find the best (MCMC-based) fit relation to be:

\begin{equation} \label{eq:mass_size}
\log(r_{e} [\rm kpc]) = 0.52\,(\pm 0.07)\,\,\log(M_{\star} [M_{\odot}])\,\, -\,\,4.98\,(\mp 0.55)
\end{equation}

Given observational limitations, it is possible that random groupings of smaller, unrelated clumps observed together due to projection effects may influence this relation. However, we argue against this scenario in Sec.~\ref{subsec:lensed_comp} and  \ref{subsec:clump_stability} based on theoretical expectations. We instead conclude that clumps at all scales are coherent, with smaller undetected sub-clumps (with masses $<10^{8.0}\,\rm M_{\odot}$ and sizes $<0.1\,\rm kpc$) being part of a hierarchy of structures.

We also examine whether including the clumps that were excluded due to a high reduced-$\chi^2$ affects our results (gray points in  Fig.~\ref{fig:clump_mass_size}).  We find that the slope of the relation then becomes $0.49\,(\pm 0.08)$, and the constant changes to $5.32\,(\mp 0.63)$. Thus, excluding clumps based on reduced-$\chi^{2}$ does not significantly change our results. Furthermore, we conclude that the excluded clumps are mainly those that are either more extended than a few kpc or smaller than $0.1\,\rm kpc$. The same arguments presented in Sec.~\ref{subsec:completeness} to explain detection limits apply here.  Clumps are challenging to model if they are much smaller than the PSF size or approach the disk size. The subsequent discussions on the mass-size relation will be taken up in Sec.~\ref{subsec:clump_stability}. 

\begin{figure} 
    \centering
    \includegraphics[width=0.49\textwidth]{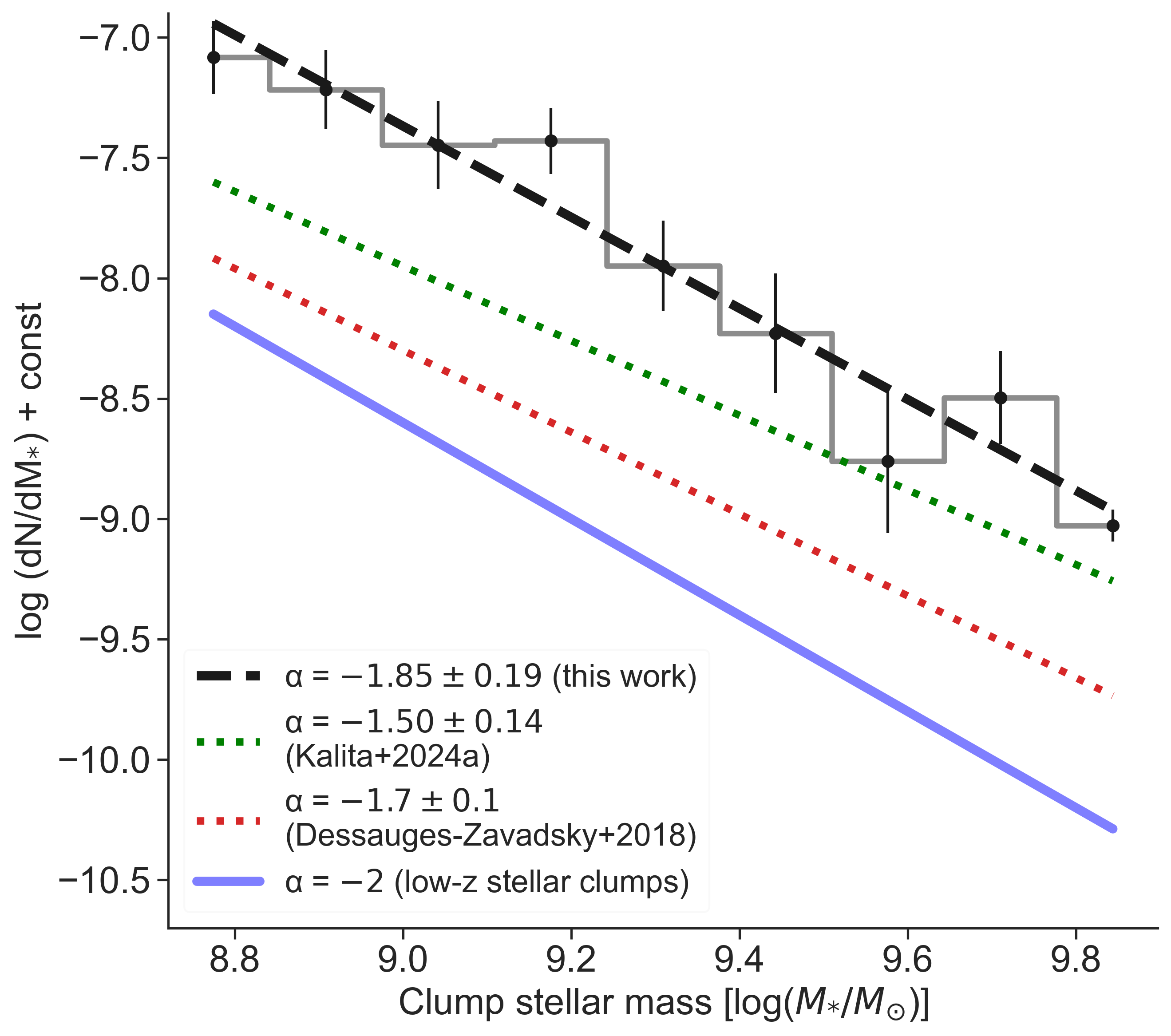}
    \caption{The clump stellar mass function in our sample is compared to previous studies \citep{kalita24b, dessauges18}. This is further compared to the theoretical prediction for clumps formed from disk instabilities \citep[$\alpha = -2$][]{elmegreen06}.  A constant has been added to the power law for a visual comparison of the slopes.}
    \label{fig:clump_mass_func}
\end{figure}

\subsection{Clump stellar mass function} \label{subsec:csmf}

The stellar mass function of clumps (cSMF) can also provide insights into their formation mechanisms \citep[e.g.,][]{elmegreen06}.  Clumps formed through turbulent instabilities leading to cascading collapse (as discussed in Sec.~\ref{subsec:clump_stability}) are expected to have a characteristic cSMF slope of $\alpha \approx -2.0 \pm 0.3$ \citep{elmegreen06,veltchev13,chandar14,adamo17,zhou24}. It is important to note, however, that these results are primarily based on low-$z$ studies, where stellar clumps are limited to masses of $\sim 10^{5}\,\rm M_{\odot}$. This upper limit is due to the physical conditions at low redshift. At $z>1$, the scales would be much larger, as argued in Sec.~\ref{subsec:clump_stability}, extending up to kpc-scale clumps with stellar masses $>10^{6}\,\rm M_{\odot}$.  Nonetheless, they would still follow the same slope, given the underlying physics remains consistent \citep{dessauges18,claeyssens24}.

It is important to note that the resolution limit may influence the fit, as the estimated mass may be a few times higher \citep{tamburello17}. However, the sensitivity limit has a more significant effect \citep{dessauges18}. Therefore, limiting our study to clumps well within our completeness range is essential for accurately measuring the clump stellar mass function. We set a limit of $10^{8.7}\,\rm M_{\odot}$, determined from the mass limit in Sec.~\ref{subsec:completeness}.  We find that the characteristic size of clumps with this mass would be $\sim 0.4\,\rm kpc$ based on Eqs.~\ref{eq:mass_size}.  Thus, we exclude clumps that are unresolved in the detection image (F150W),  without actually making it a criterion. This leaves 66 clumps in our sample.  Using the number of clumps (N) and stellar mass ($M_{\star}$), we can fit our sample with the following equation:
\begin{equation*}
\rm \log\, (dN/dM_{\star}) = \alpha\, \log\,(M_{\star}) - const 
\end{equation*}
We measure an $\alpha = -1.85 \pm 0.19$ (Fig.~\ref{fig:clump_mass_func}),  which strongly agrees with the expected $\alpha \approx -2$. We also find this slope consistent with previous studies. \cite{kalita24b} report an $\alpha = -1.50 \pm 0.14$, which is slightly shallower than the expected value due to the lower resolution resulting from the use of images PSF-matched to that of F444W.  \cite{dessauges18} with an $\alpha = -1.7 \pm 0.1$,  use clumps in lensed galaxies as well as results from an HST-based study \citep{elmegreen13} at the higher mass end.  Our measured value of $\alpha$ thus supports the conclusion that the massive stellar clumps in our study are likely part of the hierarchy of star-forming regions.

It is worth noting that we do not discuss the normalization of this relation. As highlighted by \cite{dessauges18}, normalization depends on detection thresholds and the star formation histories adopted, making direct comparisons between different studies challenging. This is why the aforementioned study does not refer to any specific value in their results. Although we find a similar normalization compared to \cite{kalita24b}, we limit our discussion to the slope, which is relatively independent of these factors.

\begin{figure} 
    \centering
    \includegraphics[width=0.48\textwidth]{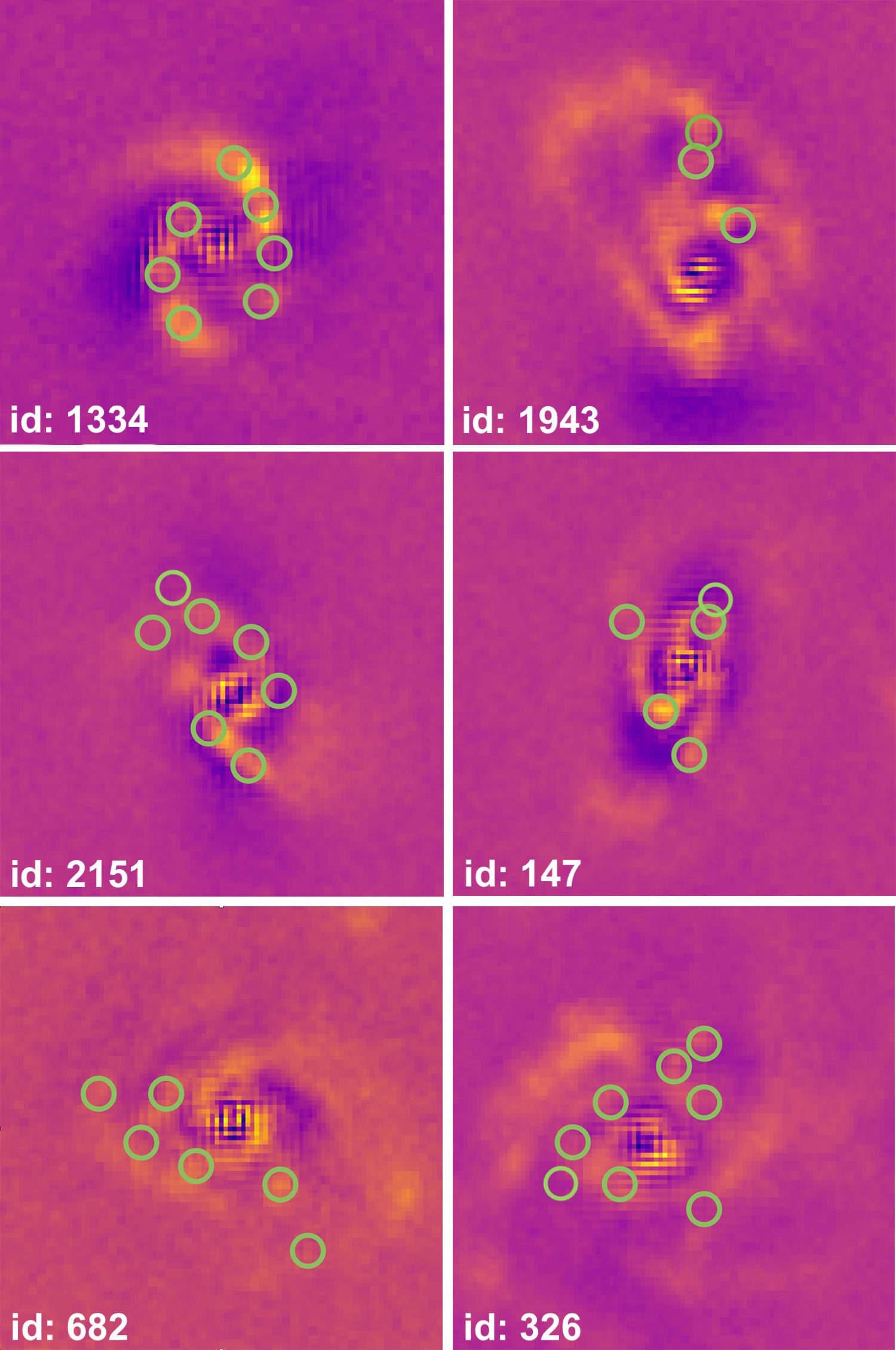}
    \caption{The location of all clumps (including detected clumps as well as those rejected due to high $\chi^2$) on the rest-frame near-IR (F444W) residual images for 6/32 galaxies in our sample. The corresponding bulge+disk+clumps model has been subtracted to produce these images.  We find $>70\%$ spatial association of clumps with residual spiral features in our study.}
    \label{fig:clumpy_residuals}
\end{figure}
\subsection{Residual spiral structures} \label{subsec:spirals_results}

A key feature of the analysis is the detection of underlying stellar structures in the form of residuals in the rest-frame near-IR images. These structures are not captured by the bulge+disk+clump models and thus appear as distinct entities in most galaxies in our study. We are not yet able to satisfactorily fit these features with models and therefore rely on visual inspection. It should be noted that galaxies with clear signs of major mergers were already removed  (Sec.~\ref{subsec:sample}).  However, based on the residuals, we find that 5 out of 26 galaxies may be undergoing detectable levels of minor mergers, which could result in tidal features being misinterpreted as spiral arms.

Interestingly,  18 of the remaining clumpy galaxies in our sample show signs of spiral arms in the rest-frame near-IR residual images, while their optical image residuals reveal these features to varying degrees of prominence. When we examine the locations of the detected clumps in these galaxies (Fig.~\ref{fig:clumpy_residuals}),  we find an overwhelming association of over $70\%$. The actual number could be higher since visual inspection can only establish associations if the residuals are detected at above $3\sigma$. Furthermore, in some cases, spatial overlap may cause some flux from the residuals to leak into the clump model. This could dampen the residual flux near the clumps, making associations more challenging to identify.

\section{Discussions}

\subsection{Multi-scale nature of clumps} \label{subsec:lensed_comp}

\begin{figure} 
    \centering
    \includegraphics[width=0.5\textwidth]{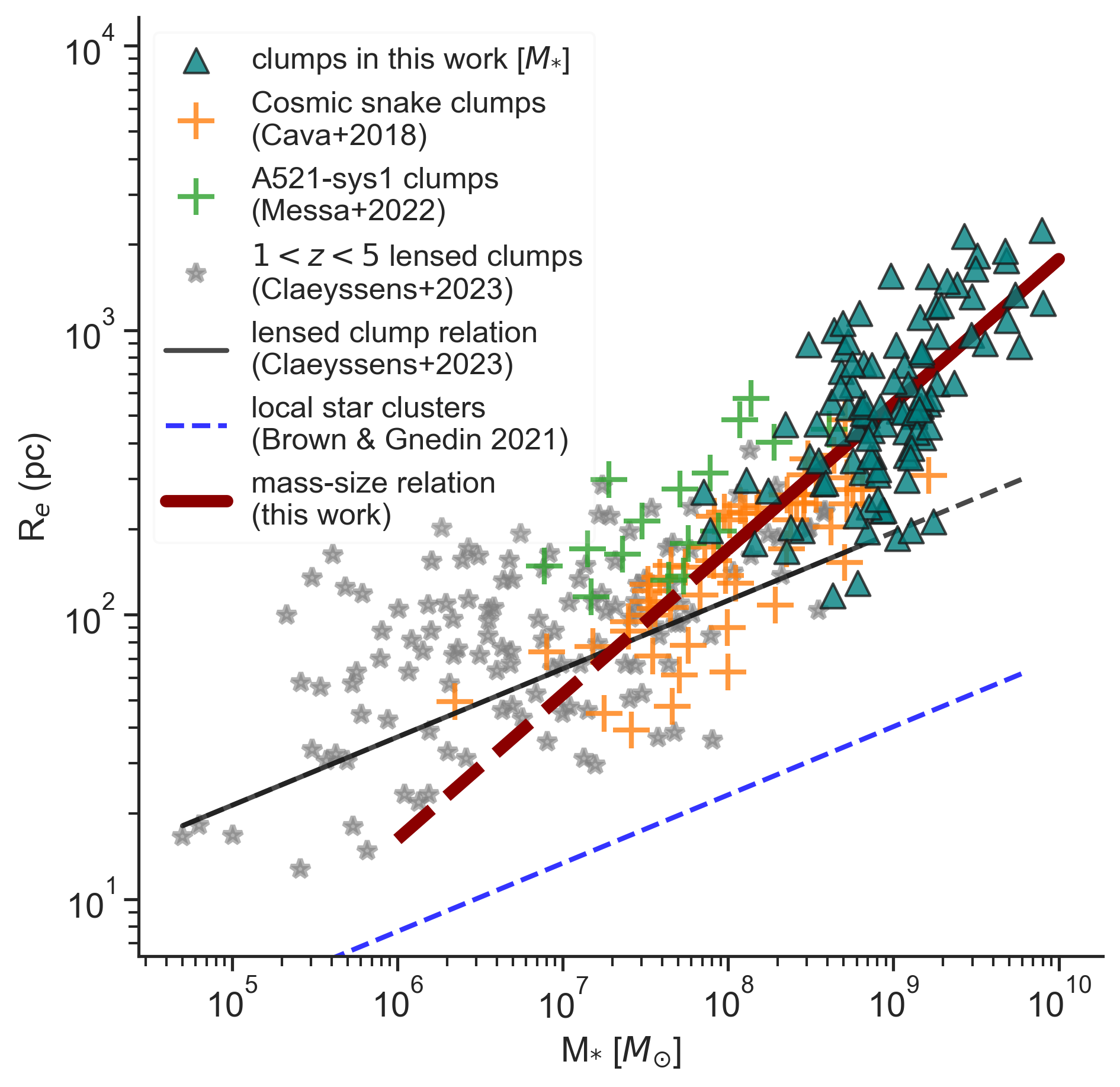}
    \caption{The comparison of the stellar clumps in this work to those in the lensed galaxies.  Clumps in the Cosmic snake \citep{cava18} and A521-sys1 \citep{messa22} are provided in orange and green.  The GMCs within these systems are later presented in Fig.~\ref{fig:gmc_comparison}.  A larger set of clumps in 11 lensed galaxies at $1 < z <5$ are shown in grey along with the corresponding scaling relation as the black dashed line \citep{claeyssens23}.  This relation also includes upper limits and clumps at $z>5$, and thus does not precisely represent the distribution of clumps in the $1 < z < 5$ range presented.  We also show the local star cluster relation \citep{brown21}. In each case,  solid lines indicate the region over which the respective relation is estimated, while dashed lines indicate extrapolation.}
    \label{fig:clump_comparison}
\end{figure}

There has been a gap in mapping the size (and hence the mass) distribution of clumps between unlensed studies \citep[investigating the massive end at $\gtrsim 1\,\rm kpc$, e.g.,][]{genzel06,elmegreen08, guo15, guo18} and lensed studies \citep[targeting structures down to tens of parsecs, e.g.,][]{jones10,livermore12,livermore15,cava18,zick20,messa22,claeyssens23,claeyssens24}. In this study, we find galaxy clumps in the range of $0.1-1\,\rm kpc$, bridging this gap and enabling more accurate size-mass measurements. This has been achieved primarily due to the high angular resolution of JWST/NIRCam images, which cover the rest-frame optical wavelengths, combined with the depth of the COSMOS-Web data.

When placed within the same mass-size parameter space, we find that the lensed clumps from 13 galaxies within $1 < z < 5$, observed over various magnifications, extend the distribution of our sample down to $\sim 10^{6}\,\rm M_{\odot}$ (Fig.~\ref{fig:clump_comparison}).  With overlap only at $\sim 100\,\rm pc$, the absence of smaller clumps in our study is due to completeness limits (Fig.~\ref{fig:completeness}).  Our sample also overlaps with other lensed systems in \cite{jones10,livermore12,livermore15}, which show similar clump sizes and stellar mass ranges of $\sim 0.06-1\,\rm kpc$ and $10^{8-9}\,\rm M_{\odot}$.  Comparisons to the (largest yet) compilation of clumps in lensed galaxies in \cite{claeyssens23,claeyssens24} also highlights the clumps in our sample overlaps with the massive end of the full range of clump masses observed in these studies.

We conclude that resolution is indeed critical in determining the detectable clump distribution. Higher resolution would likely reveal smaller clumps in both size and mass. However, this raises the question of the relevance of massive clumps. Are they merely blended star clusters observed as single objects due to resolution limits? At $\lesssim 10\,\rm pc$ scales, achievable using lensing at $z>1$, one approaches the size of star clusters 
 \citep{vanzella17a,vanzella17b,welch22,vanzella22a,vanzella22b,adamo24,messa24,mowla24}. The compilation work in \cite{claeyssens23} suggests a similar clump size-stellar mass relation ($r_{e} = 12.3 \pm 10.0\, \times \,\rm (\frac{M_{\star}}{10^{4}\,\rm M_{\odot}})^{\,0.24\pm0.10}$) to that expected in local star clusters \citep[$r_{e} = 2.55\, \times \,\rm (\frac{M_{\star}}{10^{4}\,\rm M_{\odot}})^{\,0.24}$;][also shown in Fig.~\ref{fig:clump_comparison}]{brown21}.

However, structures at larger scales ($>100\,\rm pc$) cannot be a simple extrapolation of this relation, as it would suggest an increase in mass density with size. The similarity of the clump stellar mass function (cSMF) to that of hierarchical star-forming regions at low-$z$ (Sec.~\ref{subsec:csmf}) indicates that the clump structure likely has a hierarchical nature \citep{dessauges18}. The $r_{e} \propto \rm M_{\star}^{0.52 \pm 0.07}$ relation we measure in Sec.~\ref{subsec:clump_mass_size} supports this scenario since it leads to a decreasing mass density with size. Thus, similar to low-$z$, clumps would then likely be part of hierarchical star-forming structures that are regulated by cascading turbulent motions and self-gravity \citep{elmegreen96b,elmegreen04b,bergin07,grasha17,rodriguez20}. The source of such turbulence, which influences clump formation, is debated; possible contributors include stellar feedback, gravitational energy from accreting gas, gravitational torques, and clump collisions \citep{immeli04,khochfar09,murray10,elmegreen10,ceverino10, krumholz10b}.

This hierarchical nature of clumps has been suggested by both simulations and direct observations. Several simulation studies \citep{ceverino12,behrendt16,behrendt19,faure21} find that clumps exhibit significant substructures. Observations in some cases, where different lensing magnifications reveal clumps with hierarchical structures,  also support this idea \citep{fisher17,cava18,messa19,messa22}. Therefore, if the clumps in our sample were observed with higher resolution,  one would likely detect the smaller structures within them (sub-clumps or possibly star clusters).

\subsection{Instability-driven clump formation} \label{subsec:clump_stability}

It is important to understand the physics behind the formation of these multi-scale clumps. There are two primary formation modes.  `In-situ' clumps arise as a result of disk instabilities with ages from tens (short-lived) to a few hundred Myrs (long-lived).  Meanwhile, ex-situ clumps refer to infalling stellar bodies that form outside the galaxy \citep{forster11, wuyts12, zanella15}.  The latter are expected to have characteristically higher stellar ages \citep{mandelker17},  with the assumption that these clumps do not get tidally disrupted by the galaxy potential well.  These are found to be relatively rare \citep{bournaud16},  but could contribute to the UV/Optically faint near-IR detected clumps in \cite{kalita24b}.

The age estimations from our analysis can only constrain an upper limit for the clumps at $\sim 700\,\rm Myr$ on average.  This is not sufficient to distinguish between short-lived and long-lived in-situ clumps.  However,  with the expected ages of ex-situ clumps being $\gtrsim 1\,\rm Gyr$ \citep{mandelker17},  we do not expect them to be contributing substantially to our sample.  This is likely a result of the rest-frame optical selection and a requirement for the SED to be well fit with a constant star-formation history model (Sec.~\ref{subsec:clump_prop} and Appendix Sec.~\ref{append:sed_fitting}).  The small contributution of the ex-situ clumps maybe expected to make up the most massive end of the clump mass-size relation (Fig.~\ref{fig:clump_mass_size}).  However,  visual inspection alone is insufficient to accurately determine which galaxies in our sample are undergoing minor-mergers.  A robust classification would only be possible with resolved kinematic data.

Within the instability-driven framework, clumps are believed to result from disk fragmentation \citep{elmegreen08, genzel11}. This fragmentation is governed by the Toomre instability criterion, defined as a function of disk surface mass density ($\Sigma$), epicyclic frequency ($\kappa$), and velocity dispersion ($\sigma$):
\begin{equation*}
Q = \frac{\sigma\kappa}{\pi G \Sigma} \leq 1
\end{equation*}
Based on this criterion, the Toomre length ($\lambda_{T}$) can be defined as the largest scale over which a disk can fragment due to self-gravity. As described in \cite{genzel23}, this length can be approximated by:
\begin{equation*}
\lambda_{T} = \frac{\pi^{2} G \Sigma}{v_{\rm rot}/r_{\rm disk}} \sim f_{g} \,  r_{\rm disk}
\end{equation*}
where $v_{\rm rot}$ is the disk rotational velocity, $r_{\rm disk}$ is the disk radius, and $f_{g}$ is the gas fraction. Using the values of $f_{g}$ for our sample and substituting $r_{90}$ for $r_{\rm disk}$, we find $\lambda_{T} \sim 1-5\,\rm kpc$. This indicates the largest scales below which fragmentation can occur,  which may go down to the classical Jeans length. The caveat being that the analytic Toomre theory requires $Q \approx 1$. 

Another model for instability-driven fragmentation is provided by \cite{meidt22},  where they adopt a 3D perspective of the galaxy disk.  In this new framework,  the most unstable 2D wavelength is defined as $\lambda_{2D}$.  This is the scale over which fragmentation occurs under the condition that the Toomre parameter:
\begin{equation*}
Q =  \left(\frac{\lambda_{2D}}{2~\lambda_{T}}\right)^{1/2} \leq 1
\end{equation*}
For $Q \gtrsim 1$, $\lambda_{2D}$ exceeds $\lambda_{T}$, indicating a stable disk, as rotation prevents the growth of perturbations on scales $\geq \lambda_{T}$. For $Q < 1$, $\lambda_{2D}$ can fall below $\lambda_{T}$. Assuming the galaxies in our sample are unstable with $Q < 1$, we approximate this parameter using the relation from \cite{meidt22}:
\begin{equation*}
\lambda_{2D} = \frac{2\sigma^{2}}{G\Sigma}
\end{equation*}
where $\Sigma$ is the total mass surface density of the galaxy disks\footnote{Estimated using $0.9\times$ the stellar+gas mass and $\pi \times r_{90}^{2}$.}. The parameter $\sigma$ is the velocity dispersion, assumed to approximate the ionized gas velocity dispersion. We use empirical values of $\sigma = 40 \pm 20\,\rm km/s$ for $z \sim 1.5$ from \cite{ubler19}. The estimated $\lambda_{2D}$ ranges from $\sim 0.2 - 5\,\rm kpc$, which is broadly similar to $\lambda_{T}$, but with a slightly lower limit. 

Regarding the lower limit near the Jeans length ($\lambda_{J}$), most studies ignore disk thickness while measuring it \citep[e.g.,][]{hopkins12, livermore12, livermore15}. However, considering the 3D nature of the disks, \cite{meidt22} estimates $\lambda_{J}$, which is also related to $\lambda_{2D}$:
\begin{equation*}
\lambda_{J} = \frac{\sigma\pi^{1/2}}{(G \rho)^{1/2}} = \lambda_{2D} \frac{f_{g}^{1/2}}{\pi}
\end{equation*}
where $f_{g}$ is the disk gas fraction and $\rho$ is the ISM density. As discussed in \cite{meidt23}, this can be considered the lower bound on 3D fragmentation at the mid-plane of rotating disks. We estimate $\lambda_{J}$ to be $\sim 10 - 500\,\rm pc$.  All clumps across various scales (including lensed clumps from the literature) can be hence be approximately regarded as instabilities spanning from $\lambda_{T}\,(\sim \lambda_{2D}$) down to $\sim \lambda_{J}$.  The hierarchy of the clumps discussed in Sec.~\ref{subsec:lensed_comp} therefore likely exists within this range.  Meanwhile,  the size limit of detected clumps would then depend on the resolution limit of the observations.

\subsection{GMCs and clumps}

An integral part of the hierarchical and self-similar distribution of star-forming structures in the inter-stellar medium, which we conclude the clumps to form a part of, are the gas clouds.  We therefore explore the possibility of molecular gas cloud progenitors of the massive clumps in our study.  This is especially interesting since the scaling $r_{e} \propto \rm M_{\star}^{0.52 \pm 0.07}$ for the clumps mirrors that of giant molecular clouds \citep[GMCs; size $\sim$ mass$^{0.5}$;][for a review]{heyer15}.  GMCs also exhibit a hierarchical structure shaped by turbulence, gravitational interactions, and magnetic fields, while also merging to form larger cloud complexes.  However, a direct correlation is challenging to establish since the clumps we detect are too large to form at the cores of virialized giant molecular clouds (GMCs) as observed in the local universe. Simulations also suggest that gas-rich galaxies featuring massive clumps would host GMCs of $10^{5-7}\,\rm M_{\odot}$, which are expected to be destroyed within $10-30\,\rm Myr$ by stellar feedback \citep{ceverino14,perret14}. This timescale is similar to the lifetimes of GMCs in the local Universe \citep{kawamura09,kruijssen19,chevance20,kim21}.

\begin{figure} 
    \centering
    \includegraphics[width=0.48\textwidth]{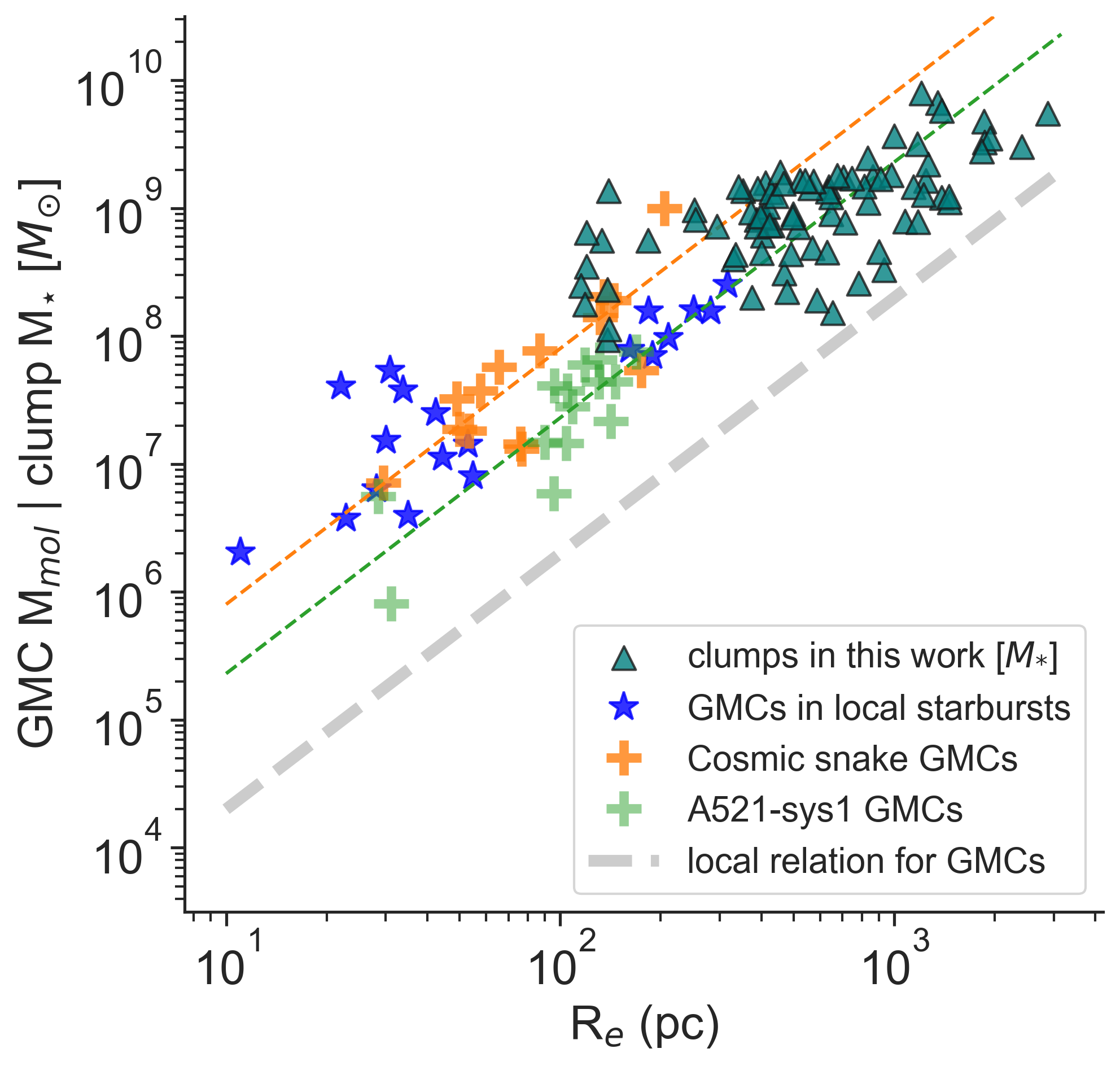}
    \caption{The comparison of the stellar clumps in this work to GMCs.  These compiled datapoints have been taken from the work that contributes the values for the A521-sys1 GMCs \citep{dessauges23}.  The other values included here are from the Cosmic snake \citep{dessauges19}, and GMCs in local starbursts \citep{wei12,leroy15}. }
    \label{fig:gmc_comparison}
\end{figure}

However, instabilities in gas disks should result in clumpy molecular gas morphologies \citep{semadini19}. Detecting such GMCs at $z \gtrsim 1$ is extremely challenging, and they have only recently been observed in a few sources \citep{dessauges19,dessauges23}. We provide a qualitative comparison of these GMCs with our clump sample in Fig.~\ref{fig:gmc_comparison}.  The distinct (but proportional) differences are also evident in the stellar clumps within these systems that host the GMCs (comparing to Fig.~\ref{fig:clump_comparison}). These stellar clumps are often spatially offset from the GMCs \citep{dessauges19,messa22}. Nevertheless,  the mass of these GMCs could represent the parent gas clouds for clumps similar to the less massive ones in these systems. It is also important to note that, at low $z$, a direct spatial association between GMCs and clumps has been observed only at scales $\gtrsim 0.5\,\rm kpc$ \citep[due to possible dissipation of GMCs by stellar feedback;][]{pan22}.

We compared these stellar clumps to our sample in Sec.~\ref{subsec:lensed_comp}.  However, it is worth noting that the two systems we refer to have gas fractions of $14-25\,\%$, while our sample has fractions of $\gtrsim 40\%$, with gas reservoirs larger by more than an order of magnitude. Consequently, the associated turbulent dispersion would result not only in larger clumps \citep{genzel11,livermore15} but also in larger GMCs \citep{larson81} in our sample. Thus, a direct one-to-one comparison would not be appropriate. More extensive GMC statistics at $z\gtrsim1$ are needed to draw a conclusion and establish an observational link between these and massive clumps.

\subsection{Clumps within spirals} \label{subsec:spirals_discussions}

The appearance of spiral arms in galaxies at $z\lesssim2.5$ \citep{margalef-bentabol22} is particularly intriguing in relation to the star-forming clumps within these systems. These "clumpy spirals" likely represent a significant phase in the evolution of massive disk galaxies, marking the transition from the high-z clumpy phase to the low-z spiral phase \citep{elmegreen14}. However, these two phases may not be mutually exclusive. Spiral arms typically form due to a low Toomre parameter ($Q < 1$) in the stellar component. Low values of $Q_{\rm star}$ result in prominent grand design spirals, while higher values of $Q_{\rm star}$ produce fainter "flocculent" spirals with patchy arms \citep{elmegreen87}. Meanwhile, the presence of massive clumps suggests low values of both $Q_{\rm star}$ and $Q_{\rm gas}$.

\cite{bournaud09b} suggests that hosts of massive clumps within spirals are likely massive galaxies with high stellar masses, which would have thick disks and even spheroids, enhancing the likelihood of spiral arm formation. Thus, it is not surprising that our sample of massive clumpy galaxies ($>10^{10.5}\,\rm M_{\odot}$) predominantly features spiral structures. However, the relative prevalence of spirals and clumps in our sample may provide further insights into the stability of these galaxies. We will explore this in follow-up papers (Kalita et al. submitted to ApJL; Kalita et al. in prep).

Finally, some regions in the rest-frame near-IR residuals may be considered clumps that remain undetected in the F150W band due to high levels of dust obscuration. A near-IR selection, as in \cite{kalita24b}, would have detected them. This further highlights the intrinsic link between clumps and extended stellar structures, with the distinction varying across wavelengths.

\section{Conclusions}

The details of clump formation are not fully understood due to limitations in depth and resolution, which typically allow the detection of only kpc-scale clumps in large statistical samples \citep[e.g.,][]{wuyts12,guo15,guo18}. In contrast, lensed studies suggest that clumps have smaller characteristic sub-kpc sizes \citep[e.g.,][]{jones10,livermore12,cava18}. Studies also indicate that the lack of high-resolution data ($\sim 100\,\rm pc$) can lead to an overestimation of physical properties \citep{tamburello15,fisher17,huertas-company20}. Proper clump characterization is therefore crucial for understanding how disk instabilities lead to their formation \citep{elmegreen09,genzel11}. JWST is well positioned to measure clump properties with high-resolution multi-wavelength data. At $z\sim1.5$, this coverage translates to the rest-frame optical to near-IR. Moreover, for the shortest wavelengths, structures like clumps can be detected down to physical scales of $\sim 0.1\,\rm kpc$.

In this work, we use data from JWST/NIRCam (4 bands) to construct bulge, disk, and clump models for 32 galaxies with spectroscopic redshift $\sim 1.5$. We also use the bulge and disk model to decompose ALMA Band-7 data. The galaxies have stellar masses above $10^{10.6}\,\rm M_{\odot}$ and lie on the star-forming main sequence. Using the rest-frame optical F150W filter, we detect and model star-forming clumps in 26 of these galaxies. The key conclusions of this paper are:
\begin{itemize}
\item The star-forming clumps have a range of stellar masses ($\sim 10^{8.1-9.5}\,\rm M_{\odot}$) and sizes ($r_{e} \sim 0.1-1\,\rm kpc$). We also assess the detection limits of our method as a function of F150W flux (and thus stellar mass) and size. The final sample of clumps is within the $90\,\%$ completeness regime. All clumps are star-forming with high $\Sigma_{\rm SFR} = 10^{-2} - 10^{2}\,\rm M_{\odot}/yr/kpc^{2}$. Each clump shows an increased sSFR compared to its host but lies within the scatter of the star-forming main sequence for its respective stellar mass.
\item We find a strong correlation between the stellar mass and size of the clumps, from which we derive an empirical stellar mass-size relation. The extension of this relation aligns with smaller clumps found in lensed studies, which may also include star clusters. Thus, we conclude that the clumps are part of a hierarchical structure, similar to that observed in star-forming regions in the local Universe.
\item We measure a stellar mass function slope of $-1.85 \pm 0.19$ for the clumps, in remarkable agreement with the hierarchical star-forming regions ($\alpha \approx -2$) observed in local galaxies. This also suggests that these structures are likely shaped by multi-scale turbulent motions and self-gravity.
\item We conclude that the clumps detected in our study along with those which maybe below our detection limit ($<100\,\rm pc$) likely cover the range of scales bound by the Toomre length ($\sim$ a few $\rm kpc$) and the Jeans length ($\sim$ tens to a few hundred $\rm pc$).
\item After subtracting the bulge, disk, and clump components, we clearly observe spiral features in the rest-frame near-IR residual images. There is an association of $>70\%$ between the clumps and the spiral features. We require to properly characterize these features to help us understand the evolution of the dynamical stability of massive stellar disks at high $z$.
\end{itemize}

In conclusion, we find that the improved imaging using JWST/NIRCam and modeling capabilities of this study allow us to properly characterize clumps in the most massive star-forming galaxies at $z\sim1.5$. The observational divide between lensed and unlensed studies is reduced, thereby distributing their properties over a wide range of characteristic mass and size. We find that the entire clump population can be explained by instability-driven coherent collapse and resulting star formation. The limitations of using only imaging data and, consequently, anticipating the dynamics involved can be overcome by obtaining high-resolution spectroscopic data. This will provide the final piece of this coherent picture of galaxy clumps.

\section*{Acknowledgements}

We thank the referee for the crucial comments and suggestions. B.S.K. and J.D.S. are supported by the World Premier International Research Center Initiative (WPI), MEXT, Japan. J.D.S. is supported by JSPS KAKENHI (JP22H01262). This work was also supported by JSPS Core-to-Core Program (grant number: JPJSCCA20210003). W.M.  and O.I.  acknowledge the funding of the French Agence Nationale de la Recherche for the project iMAGE (grant ANR-22-CE31-0007).  L.C.H. was supported by the National Science Foundation of China (11991052, 12233001), the National Key R\&D Program of China (2022YFF0503401), and the China Manned Space Project (CMS-CSST-2021-A04, CMS-CSST-2021-A06).  This work was made possible by utilising the CANDIDE cluster at the Institut d’Astrophysique de Paris. The cluster was funded through grants from the PNCG, CNES, DIM-ACAV, the Euclid Consortium, and the Danish National Research Foundation Cosmic Dawn Center (DNRF140). It is maintained by Stephane Rouberol. We would like to thank Miroslava Dessauges-Zavadsky and Sharon Meidt for the valuable discussions which helped shape the scientific narrative of this work.

\section*{Data Availability}

All JWST and HST data products are available via the Mikulski Archive for Space Telescopes (https://mast.stsci.edu). The  ALMA data are available on the ALMA Science Archive at https://almascience.eso.org/asax/.



\bibliographystyle{mnras}
\bibliography{main} 




\appendix

\section{Spatial Deconstruction: JWST} \label{append:jwst_deconstruction}

\begin{figure*} 
    \centering
    \includegraphics[width=0.95\textwidth]{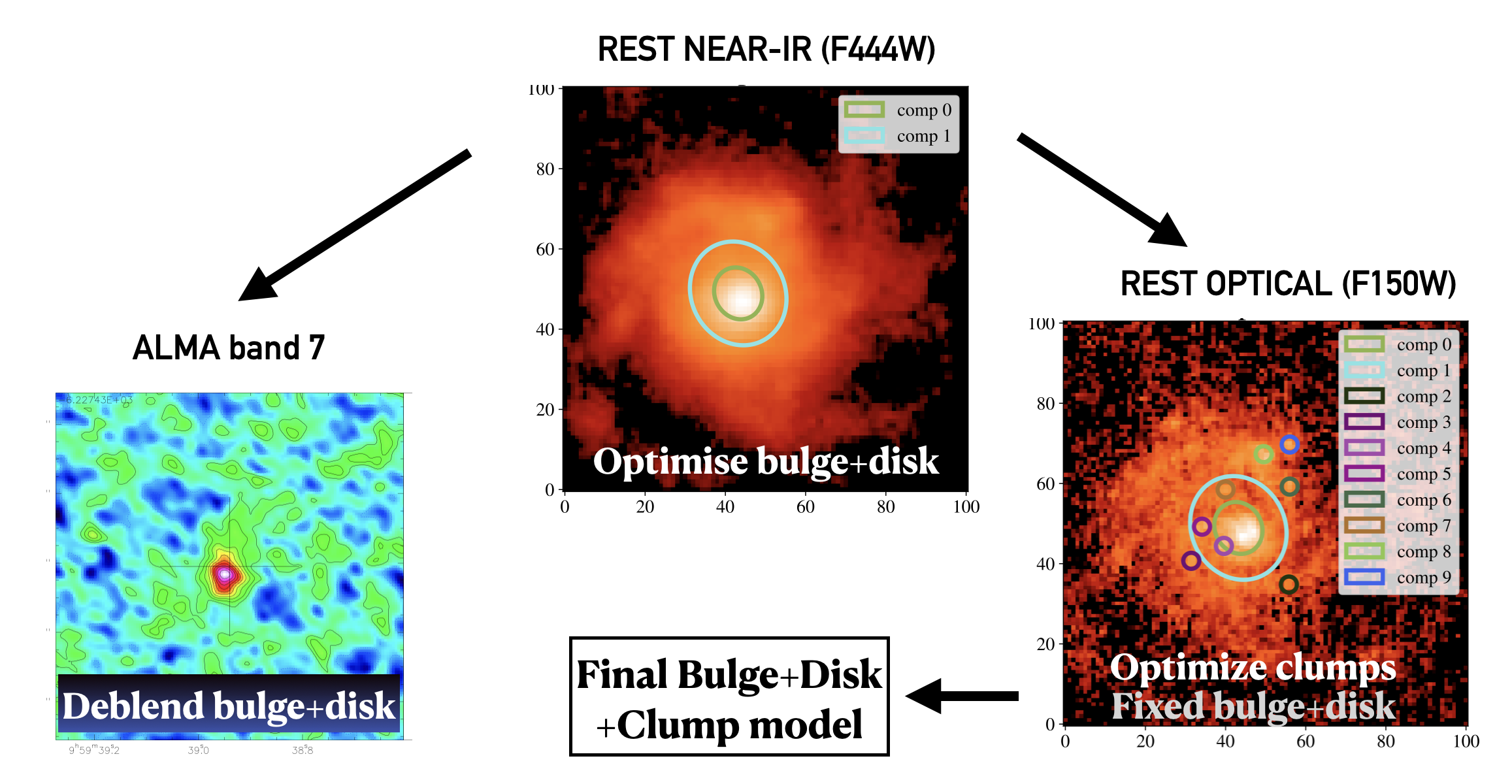}
    \caption{A flowchart describing the process of galaxy deconstruction used in this work.  We start with optimising the bulge and disk model in F444W.   This is used along with clumps in the F150W image to finally get a bulge+disk+clump model.  The ellipses indicate the \emph{initial} guesses for each component.  The same bulge and disk model is used to deblend the flux in the ALMA band-7 image to divide to net flux into a bulge and disk component. }
    \label{fig:flowchart}
\end{figure*} 

Here we detail the spatial `deconstruction' of JWST images of galaxies into their constituent components: bulge, disk, and clumps (Fig.~\ref{fig:flowchart}). This was briefly discussed in Sec.~\ref{subsec:jwst_measurements}.  For this method, we use the rest-frame near-IR F444W image to optimize the bulge+disk model and the rest-frame optical F150W image to optimize the star-forming clumps. The fitting is performed using the Python-based package \textsc{GALIGHT}\footnote{https://github.com/dartoon/galight} \citep{ding22}, which implements the forward-modeling galaxy image fitting tool \textsc{LENSTRONOMY}\footnote{https://github.com/lenstronomy/lenstronomy} \citep{birrer18,birrer21}. This approach provides access to the full posterior distribution of each fitted parameter. The fitting is then optimized using the Particle Swarm Optimizer \citep[PSO]{kennedy95}. All errors are estimated by artificially adding structures of similar sizes and fluxes and remeasuring them (Tang et al. in prep). The PSFs are created using the software \textsc{PSFEx} \citep{bertin11} on the full COSMOS-Web mosaics \citep{casey23}.

\subsection{Bulge and disk in F444W} \label{ap:bulge_disk}

To fit the F444W image, we start with a single Sérsic fit, allowing all spatial parameters to vary freely, as a reference. Given that the galaxies in our sample lie on the main sequence and exhibit centrally concentrated star formation (Suzuki et al., in prep.), we use a fixed Sérsic index ($n=2$) for the bulge and $n=1$ for the disk, assuming a pseudo-bulge rather than a classical bulge ($n=4$). However, we find that our results remain unchanged even if $n=4$ is used. In each of the 32 galaxies in our final sample, the bulge and disk model provides a better fit than the single Sérsic fit ($\rm BIC_{bulge+disk} < BIC_{single\_sersic}$). With the final bulge+disk model, we fix all shape parameters and allow only the flux of the bulge to vary independently when fitting the rest-frame optical (F150W) image. As discussed in the main text, we find very high BIC values and significant residual flux, indicating the need for additional models to account for the clumpy substructures, which will be addressed in the next subsection.

\subsection{Adding the clumps in F150W}

The clumps with high levels of $\Sigma_{\rm SFR}$ (Sec.~\ref{subsec:clump_prop}),  are most prominent in the rest-frame optical (Fig.~\ref{fig:clump_detectability}).  We therefore use the F150W image to model the clump profiles. We start with the widely used detection method \citep[most recently in][]{kalita24}, where a contrast map is created using the difference between the original image and a smoothed version (with a Gaussian kernel of $\sigma = 3$ pixels). Using a broader kernel for smoothing would prevent detection of the smallest sources, limiting the resolution of the F150W. Any smaller smoothing will approach the PSF, decreasing the flux in the contrast image and making clump detection more challenging.

We then apply $\sigma$-thresholding to detect clumps in the residual image, estimating $\sigma$ using the \emph{sigma\_clipped\_stats} feature in the \textsc{astropy} package \citep{astropy22}. Starting with a depth of $10\,\sigma$ to identify clump locations, we progressively lower the threshold to $3\,\sigma$.  Any detection with $<5$ pixels will not be considered. At each step, we use the bulge+disk model from F444W and add Gaussian profiles at the clump locations (initialized with sizes equal to 1.5 times the PSF), allowing the clump flux and size to vary. The shape of the bulge+disk model is fixed, but the flux of each component is allowed to change. We estimate the BIC at each step to assess the goodness of fit given the model's added complexity. In each case, the best BIC is reached at either $3$ or $4\,\sigma$. At the end of this process, we obtain a final bulge+disk+clump model, which we use with fixed shapes to measure the flux of each component in each filter.
\begin{figure} 
    \centering
    \includegraphics[width=0.5\textwidth]{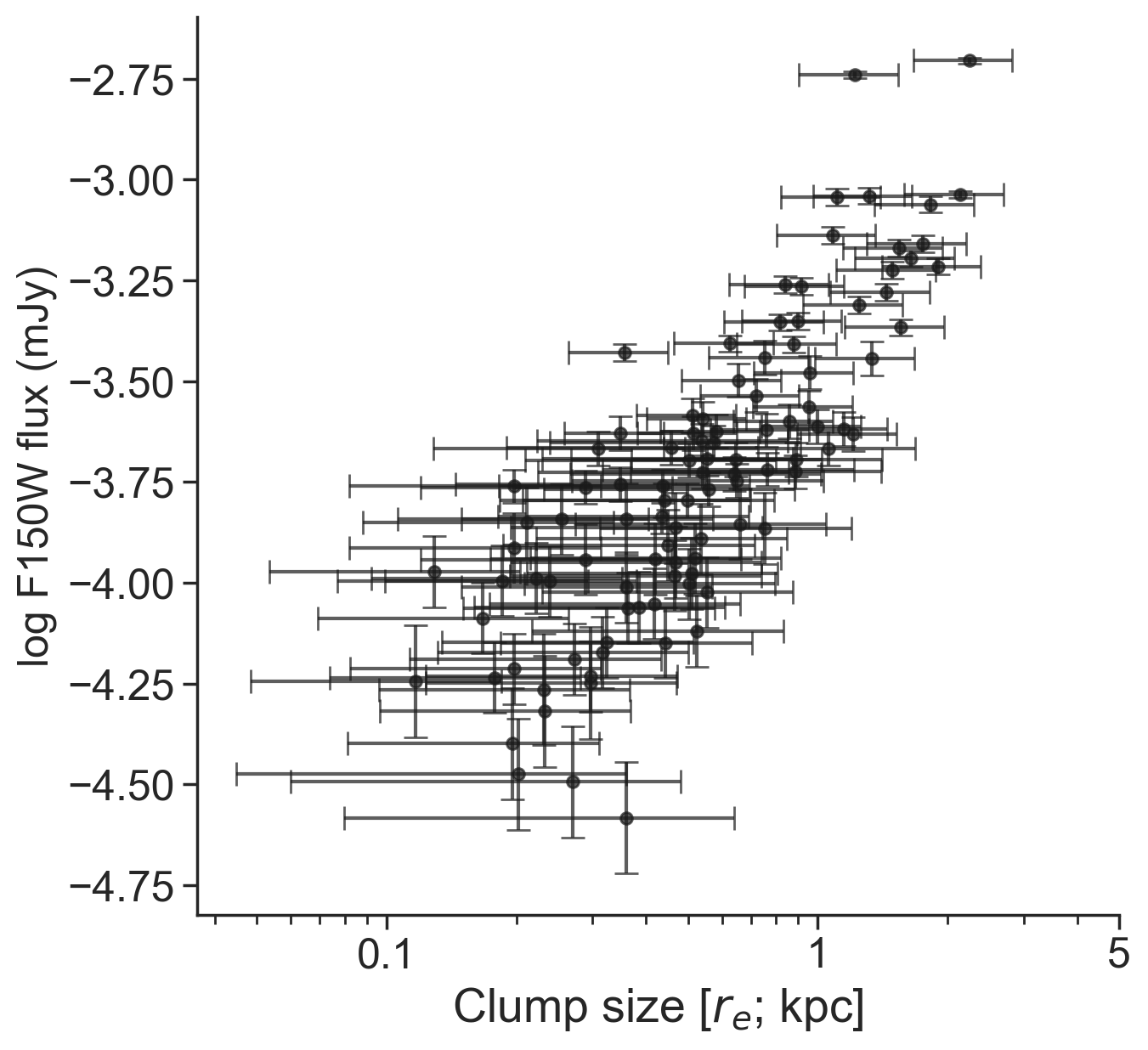}
    \caption{The clump size ($r_{e}$) vs the F150W flux for the clumps in our sample.}
    \label{fig:size_and_light}
\end{figure}

\section{Spatial Deconstruction: ALMA band-7} \label{ap:alma_meas}

For the ALMA Band-7 flux measurements, we use only the bulge and disk model, as the resolution and signal-to-noise are insufficient to measure the clumps. We assume that the $870,\mu$m flux can be decomposed into bulge and disk components, based on recent observational evidence that the sub-mm flux follows the stellar mass maps traced by the rest-frame near-IR.

For the fitting, we use the UV-plane rather than the image plane, which has been shown to perform better due to the absence of mathematical approximations involved in deconvolution and imaging \citep{tan24}. The software used is \textsc{GILDAS}\footnote{\url{http://www.iram.fr/IRAMFR/GILDAS}}, which has been widely employed for complex morphological modeling of ALMA images in several studies \citep[e.g.,][]{kalita22,tan24}.

Since the Sérsic profile used in the bulge+disk model cannot be directly converted to the UV-plane due to its Fourier transform not being analytically expressible, we instead use the Spergel profile \citep{spergel10}. This profile correlates well with the Sérsic profile, allowing us to translate the bulge and disk profiles from F444W into the UV-plane. This process also involves using a Spergel index $\nu$, empirically determined as a function of the Sérsic index and effective radius \citep{tan24}. We then deblend the ALMA Band-7 flux by fixing the shape parameters (Fig.~\ref{fig:flowchart}).

\section{SED fitting using CIGALE} \label{append:sed_fitting}

\begin{figure*} 
    \centering
    \includegraphics[width=0.48\textwidth]{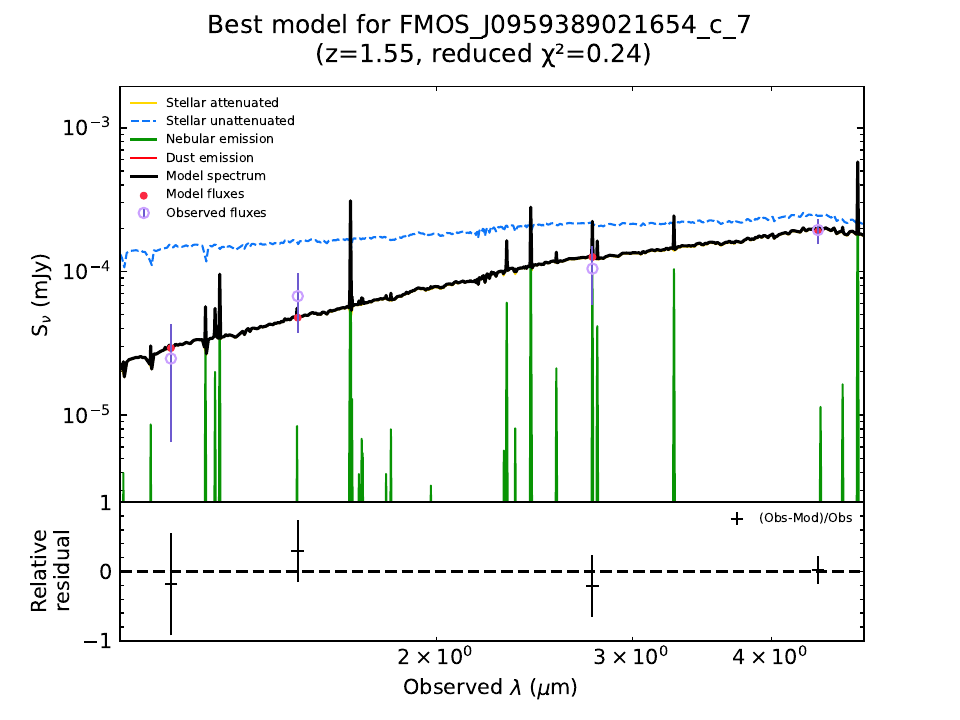}
    \includegraphics[width=0.48\textwidth]{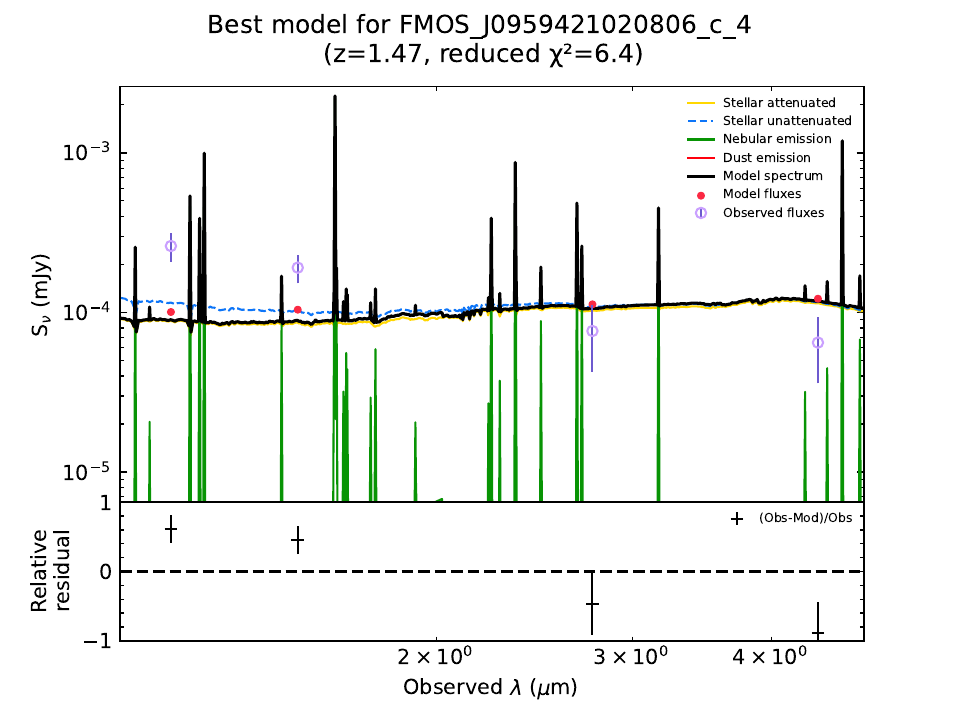}
    \caption{SED fit example of a fit for a clump that was included (left) and another that was excluded (right) in our sample. }
    \label{fig:SED_fits_clumps}
\end{figure*}

\begin{figure*} 
    \centering
    \includegraphics[width=0.48\textwidth]{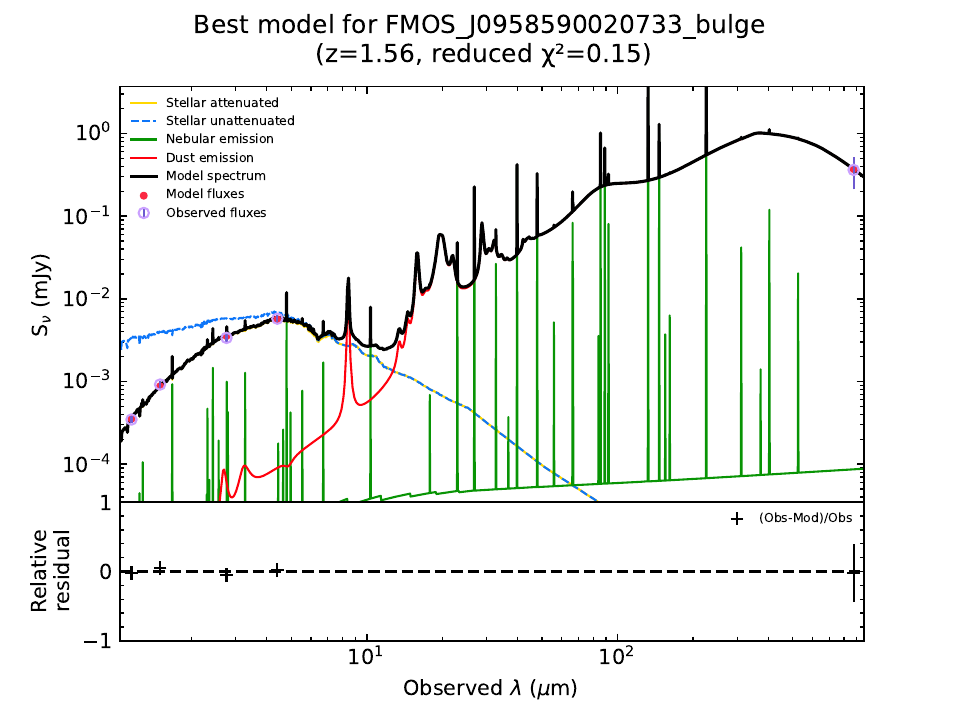}
    \includegraphics[width=0.48\textwidth]{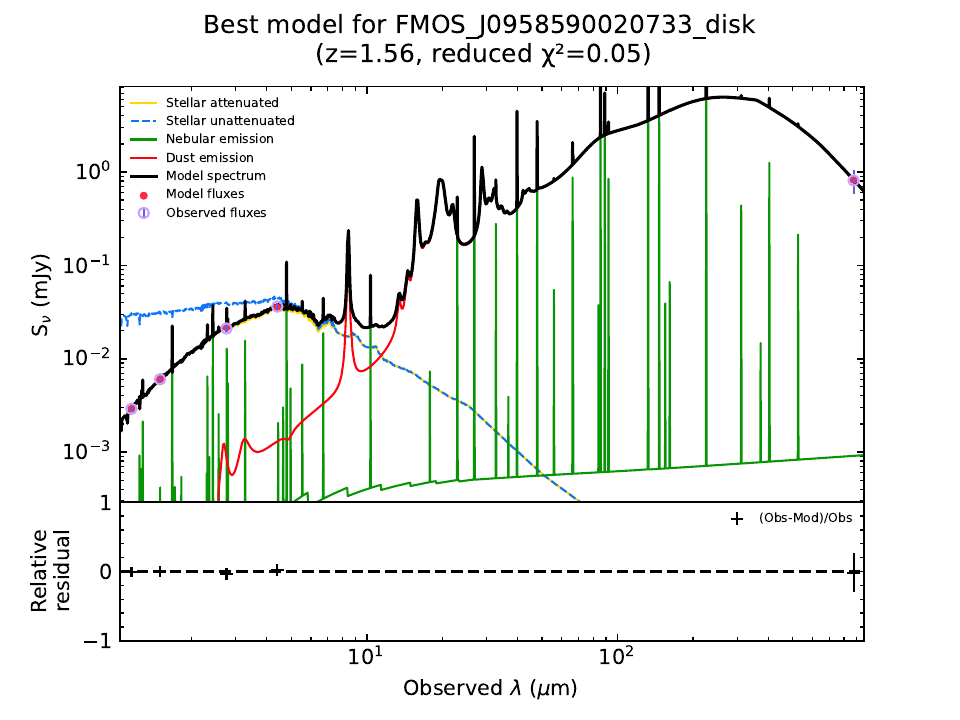}
    \caption{SED fit example with 4 JWST and 1 ALMA bands of a fit for bulge (left) and disk (right) in source id 147.}
    \label{fig:SED_fits_bulge_disk}
\end{figure*}
\begin{figure} 
    \centering
    \includegraphics[width=0.5\textwidth]{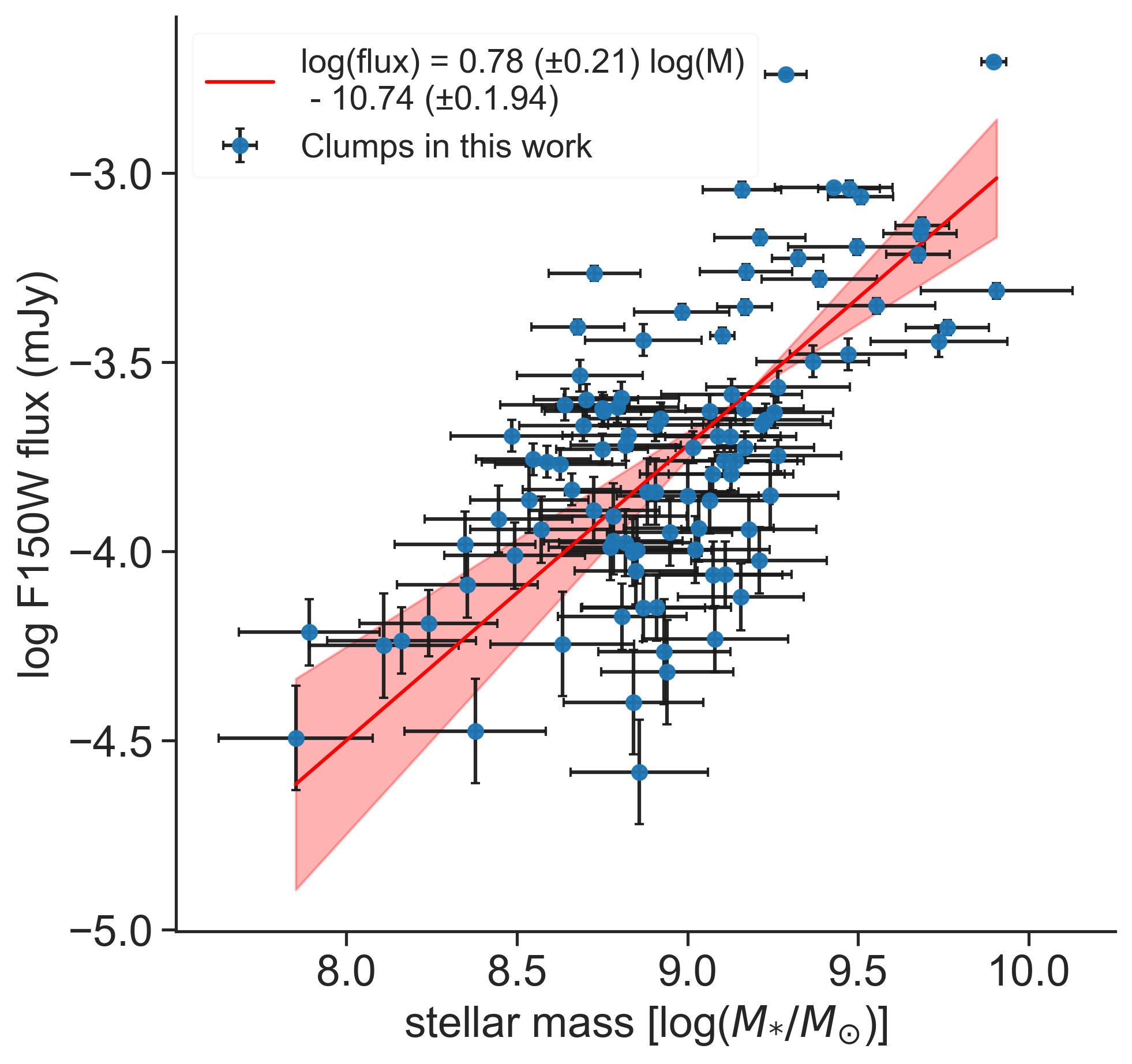}
    \caption{The mass-to-light relation for the clumps in our sample,  which relates the F150W flux to the SED-based stellar mass determination.}
    \label{fig:mass_to_light}
\end{figure}
The fluxes measured in the 4 JWST bands and 1 ALMA band are used in the SED fitting procedure, utilizing the python-based ``Code Investigating GALaxy Emission" \citep[CIGALE;][]{boquien19} tool. CIGALE relies on the energy balance between stellar emission and dust re-emission. Our fitting procedure is similar to that of Kashino et al. (in prep), but simpler in some aspects to avoid over-complication due to the limited number of flux measurements. The key details are as follows:

\begin{itemize}[leftmargin=*, noitemsep, topsep=0pt]
\item Star-formation history (SFH): Given that the galaxies are selected to be star-forming, with clumps particularly active, we adopt a constant SFR with varying duration (age of the main stellar population). 
\item Single Stellar Population (SSP): We use the widely adopted \cite{bruzual03} stellar population synthesis model with the \cite{chabrier03} initial mass function, accounting for nebular continuum and line emission. 
\item No AGN template: Any AGN hosts, identified using various tracers including X-ray, mid-IR, and broad-line components in $\rm H\alpha$, have already been excluded. Additionally, we observe no difference in the measured properties of the bulges and disks when using an AGN template. Thus, we do not use one for our final run. 
\item Dust: We apply a modified Calzetti law, based on the original model \cite{calzetti00} extended between the Lyman break and $150,\rm nm$ \citep{leitherer02}, improving applicability to star-forming galaxies at higher redshifts. We use the \cite{draine07} dust emission models to represent the sub-mm flux and the radiation field intensity within the range $U_{\rm min} = 0.1$ to $U_{\rm max} = 10^{6}$. 
\item Molecular gas mass: The dust masses from the SED fitting are converted to molecular gas mass using the metallicity-dependent gas-to-dust ratio \citep{magdis12}, based on the [NII]/$\rm H\alpha$ ratio \citep{kashino19}.
\end{itemize}

\begin{figure*} 
    \centering
    \includegraphics[width=0.95\textwidth]{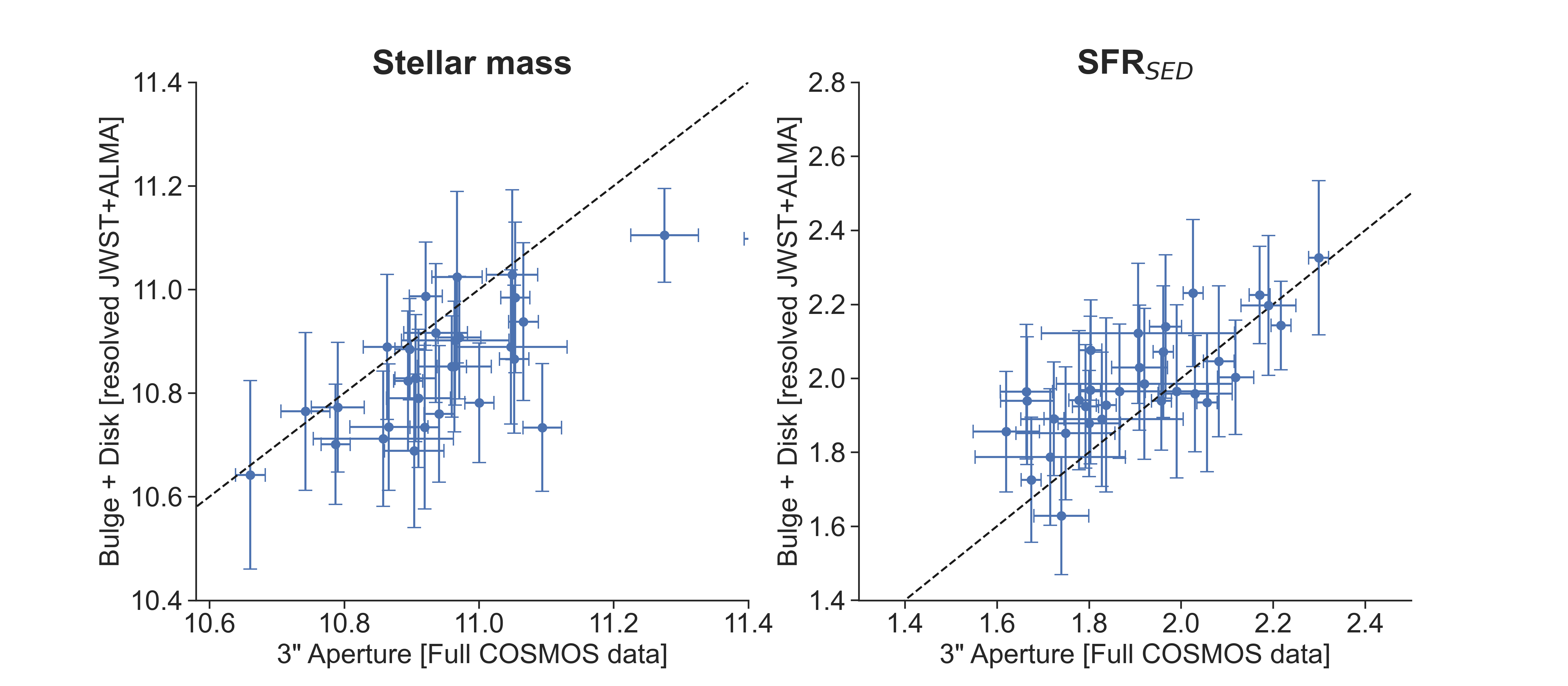}
    \caption{Comparison of the sum of log stellar mass and log SFR measurements for the bulge and the disk,  to the results from the $3^{\prime\prime}$ aperture flux measurements of the whole galaxy.}
    \label{fig:kashino_comp}
\end{figure*}

For the bulge and disk measurements, we use the full 5-band photometry. In the JWST bands, the disk flux is the sum\footnote{with the net error the quadrature sum of individual components} of the disk model flux and the flux of clumps within $r_{90}$ of the rest-frame near-IR disk. However, for the clump SED fitting, we do not have access to the ALMA Band-7 fluxes and thus perform only a 4-band SED fitting. For both, we use the $\rm H\alpha$-determined spectroscopic redshift. We provide the final results in Table.~\ref{tab:1}.  We also compare the sum of stellar mass and SFR measurements for the bulge and disk to those from $3^{\prime\prime}$ aperture flux measurements encompassing the complete host (discussed in Sec.~\ref{subsec:completeness}) in Fig.~\ref{fig:kashino_comp}.  The respective error bars reflect the uncertainties of the measurements (derived from values at $\Delta \chi^{2} = \pm 1$) in the two works.

\begin{table*}
\centering
\caption{The details of the final 32 galaxies in our sample (Fig.~\ref{fig:image_comp}). The galaxy stellar masses (Column 5) have been measured using the COSMOS2020 flux catalogue \citep{weaver22} and the same SED fitting recipe used in this work.  The last column provides the number of detected clumps in each galaxy that has been included in the final sample.  The corresponding number of clumps before rejecting based on $\chi^2$ is also provided within brackets.}
\label{tab:1}
\begin{tabular}{|c|c|c|c|c|c|c|c|c|c|}
\hline\hline
ID & RA & Dec & Spec-z & Stellar Mass & Bulge St.  mass & Bulge SFR & Disk St.  mass & Disk SFR & Clump count\\
 & (deg) & (deg) &  & $\rm \log (M_{\star}/M_{\odot})$ & $\rm \log (M_{\star}/M_{\odot})$ & $\rm \log (M_{\odot}/yr)$ & $\rm \log (M_{\star}/M_{\odot})$ & $\rm \log (M_{\odot}/yr)$ & \\
\hline
1987 & 149.721750 & 1.965500 & 1.626438 & 10.97 ± 0.04 & 10.25 ± 0.21 & 2.15 ± 0.22 & 10.95 ± 0.20 & 2.15 ± 0.22 & 0 (4)\\
147 & 149.745833 & 2.125944 & 1.556338 & 11.07 ± 0.02 & 10.07 ± 0.18 & 2.00 ± 0.20 & 10.88 ± 0.17 & 2.00 ± 0.20 & 3 (5)\\
1704 & 149.775542 & 2.251639 & 1.673098 & 10.90 ± 0.04 & 9.73 ± 0.17 & 1.83 ± 0.20 & 10.65 ± 0.16 & 1.83 ± 0.20 & 0 (1)\\
2423 & 149.825125 & 1.978889 & 1.487085 & 10.91 ± 0.03 & 10.60 ± 0.17 & 1.85 ± 0.22 & 10.45 ± 0.16 & 1.85 ± 0.22 & 1 (8)\\
1943 & 149.832750 & 2.024306 & 1.482637 & 10.90 ± 0.02 & 10.06 ± 0.20 & 1.76 ± 0.17 & 10.82 ± 0.11 & 1.76 ± 0.17 & 1 (3)\\
1147 & 149.854708 & 2.115111 & 1.481732 & 11.05 ± 0.08 & 10.34 ± 0.17 & 2.01 ± 0.24 & 10.74 ± 0.20 & 2.01 ± 0.24 & 0 (2)\\
89 & 149.855750 & 2.130222 & 1.478401 & 11.09 ± 0.03 & 10.27 ± 0.13 & 1.74 ± 0.23 & 10.55 ± 0.18 & 1.74 ± 0.23 & 2 (2)\\
1831 & 149.877875 & 2.297639 & 1.509195 & 10.96 ± 0.05 & 10.32 ± 0.15 & 1.90 ± 0.21 & 10.71 ± 0.16 & 1.90 ± 0.21 & 0 (3)\\
1766 & 149.878792 & 2.499833 & 1.672921 & 10.92 ± 0.02 & 10.36 ± 0.20 & 2.01 ± 0.17 & 10.87 ± 0.13 & 2.01 ± 0.17 & 3 (3)\\
682 & 149.889708 & 2.358778 & 1.501451 & 10.79 ± 0.02 & 9.96 ± 0.16 & 1.63 ± 0.20 & 10.61 ± 0.14 & 1.63 ± 0.20 & 2 (6)\\
326 & 149.912250 & 2.281639 & 1.550443 & 10.96 ± 0.02 & 10.40 ± 0.20 & 1.65 ± 0.16 & 10.64 ± 0.11 & 1.65 ± 0.16 & 7 (8)\\
2028 & 149.923208 & 1.898194 & 1.552804 & 11.05 ± 0.02 & 9.91 ± 0.18 & 1.89 ± 0.21 & 10.81 ± 0.16 & 1.89 ± 0.21 & 0 (2)\\
193 & 149.925333 & 2.135028 & 1.470131 & 10.91 ± 0.05 & 10.14 ± 0.17 & 1.85 ± 0.29 & 10.66 ± 0.18 & 1.85 ± 0.29 & 2 (6)\\
1945 & 149.938375 & 1.950056 & 1.444649 & 10.88 ± 0.05 & 9.98 ± 0.23 & 1.19 ± 0.22 & 10.15 ± 0.15 & 1.19 ± 0.22 & 1 (3)\\
285 & 149.972667 & 2.490139 & 1.455462 & 10.92 ± 0.02 & 9.93 ± 0.18 & 1.90 ± 0.24 & 10.65 ± 0.19 & 1.90 ± 0.24 & 0 (1)\\
3074 & 149.985375 & 2.561778 & 1.557741 & 10.96 ± 0.08 & 10.23 ± 0.12 & 1.90 ± 0.22 & 10.80 ± 0.16 & 1.90 ± 0.22 & 1 (2)\\
2074 & 149.997875 & 2.480861 & 1.462140 & 11.05 ± 0.02 & 10.22 ± 0.17 & 2.07 ± 0.23 & 10.90 ± 0.17 & 2.07 ± 0.23 & 14 (15)\\
1861 & 150.001083 & 2.321444 & 1.459471 & 11.00 ± 0.02 & 10.34 ± 0.17 & 1.69 ± 0.21 & 10.59 ± 0.15 & 1.69 ± 0.21 & 2 (3)\\
1749 & 150.025000 & 2.355278 & 1.610989 & 10.87 ± 0.06 & 10.30 ± 0.20 & 1.72 ± 0.23 & 10.53 ± 0.18 & 1.72 ± 0.23 & 2 (3)\\
281 & 150.124042 & 2.447583 & 1.599120 & 10.97 ± 0.03 & 10.61 ± 0.16 & 1.78 ± 0.22 & 10.59 ± 0.18 & 1.78 ± 0.22 & 3 (3)\\
81 & 150.164458 & 1.936306 & 1.525303 & 10.94 ± 0.02 & 10.09 ± 0.19 & 1.84 ± 0.21 & 10.65 ± 0.16 & 1.84 ± 0.21 & 7 (8)\\
2151 & 150.200875 & 2.460639 & 1.582532 & 10.79 ± 0.04 & 10.22 ± 0.20 & 1.70 ± 0.23 & 10.63 ± 0.16 & 1.70 ± 0.23 & 4 (7)\\
1663 & 150.220708 & 2.013139 & 1.604375 & 10.86 ± 0.04 & 10.11 ± 0.17 & 1.99 ± 0.23 & 10.81 ± 0.17 & 1.99 ± 0.23 & 4 (4)\\
482 & 150.258083 & 2.240167 & 1.603783 & 10.89 ± 0.02 & 10.15 ± 0.13 & 1.88 ± 0.22 & 10.72 ± 0.17 & 1.88 ± 0.22 & 7 (7)\\
428 & 150.292500 & 2.422167 & 1.635788 & 11.05 ± 0.04 & 10.19 ± 0.15 & 2.19 ± 0.22 & 10.96 ± 0.19 & 2.19 ± 0.22 & 7 (11)\\
2905 & 150.302250 & 1.931889 & 1.548028 & 10.94 ± 0.05 & 9.89 ± 0.17 & 1.86 ± 0.19 & 10.87 ± 0.14 & 1.86 ± 0.19 & 4 (7)\\
1394 & 150.365458 & 2.227500 & 1.699351 & 11.28 ± 0.05 & 10.38 ± 0.17 & 1.99 ± 0.14 & 11.02 ± 0.11 & 1.99 ± 0.14 & 3 (3)\\
3165 & 150.394917 & 2.456361 & 1.438470 & 10.74 ± 0.04 & 10.12 ± 0.17 & 1.91 ± 0.24 & 10.65 ± 0.19 & 1.91 ± 0.24 & 6 (9)\\
1334 & 150.402917 & 2.408833 & 1.514096 & 10.54 ± 0.04 & 9.93 ± 0.16 & 1.88 ± 0.20 & 10.79 ± 0.15 & 1.88 ± 0.20 & 3 (7)\\
852 & 150.441542 & 2.129222 & 1.557442 & 11.42 ± 0.02 & 10.44 ± 0.12 & 2.27 ± 0.23 & 10.98 ± 0.22 & 2.27 ± 0.23 & 6 (10)\\
4003 & 150.469292 & 2.476528 & 1.579673 & 10.86 ± 0.10 & 10.00 ± 0.15 & 1.69 ± 0.22 & 10.62 ± 0.16 & 1.69 ± 0.22 & 4 (4)\\
495 & 150.482708 & 2.304139 & 1.485021 & 10.66 ± 0.02 & 9.68 ± 0.18 & 1.90 ± 0.23 & 10.59 ± 0.18 & 1.90 ± 0.23 & 3 (7)\\
\hline
\end{tabular}
\end{table*}

\vspace{5in}
\newpage
\noindent
\textit{\footnotesize
$^{1}$Kavli IPMU (WPI), UTIAS, The University of Tokyo, Kashiwa, Chiba 277-8583, Japan \\
$^{2}$Kavli Institute for Astronomy and Astrophysics, Peking University, Beijing 100871, People{\textquotesingle}s Republic of China\\
$^{3}$Centre for Data-Driven Discovery, Kavli IPMU (WPI), UTIAS, The University of Tokyo, Kashiwa, Chiba 277-8583, Japan\\
$^{4}$National Astronomical Observatory of Japan, 2-21-1 Osawa, Mitaka, Tokyo 181-8588, Japan\\
$^{5}$Department of Astronomy, School of Science, The University of Tokyo, 7-3-1 Hongo, Bunkyo, Tokyo 113-0033, Japan\\
$^{6}$Centre for Astrophysical Sciences, Department of Physics \& Astronomy, Johns Hopkins University, Baltimore, MD 21218, USA\\
$^{7}$Universit{\'e} Paris-Saclay, Universit{\'e} Paris Cit{\'e}, CEA, CNRS, AIM, Paris, 91191,France\\
$^{8}$Department of Astronomy, School of Physics, Peking University, Beijing 100871, People{\textquotesingle}s Republic of China\\
$^{9}$School of Physics and Technology, Wuhan University, Wuhan 430072,  China\\
$^{10}$Aix Marseille Univ, CNRS, CNES, LAM, Marseille, France\\
$^{11}$Caltech/IPAC, MS 314-6, 1200 E. California Blvd. Pasadena, CA 91125, USA\\
$^{12}$NASA Headquarters, 300 Hidden Figures Way, SE, Mary W. Jackson NASA HQ Building, Washington, DC 20546, USA\\
$^{13}$Cosmic Dawn Center (DAWN), Denmark\\
$^{14}$Niels Bohr Institute, University of Copenhagen, Jagtvej 128, DK-2200 Copenhagen N, Denmark\\
$^{15}$School of Physics and Astronomy, University of Southampton, Highfield SO17 1BJ, UK\\
$^{16}$Faculty of Global Interdisciplinary Science and Innovation, Shizuoka University, 836 Ohya, Suruga-ku, Shizuoka 422-8529, Japan\\
$^{17}$Centre for Astrophysics Research, University of Hertfordshire, College Lane, Hatfield AL10 9AB, UK.\\
$^{18}$Laboratory for Multiwavelength Astrophysics, School of Physics and Astronomy, Rochester Institute of Technology, 84 Lomb Memorial Drive, Rochester, NY 14623, USA\\
$^{19}$Astrophysics Division, NASA Goddard Space Flight Center, Greenbelt, MD, 20771, USA\\
$^{20}$Department of Astronomy, School of Science, The University of Tokyo, 7-3-1 Hongo, Bunkyo, Tokyo 113-0033, Japan\\
$^{21}$DTU-Space, Technical University of Denmark, Elektrovej 327, 2800, Kgs. Lyngby, Denmark\\
$^{22}$Purple Mountain Observatory \& Key Laboratory for Radio Astronomy, Chinese Academy of Sciences, 10 Yuanhua Road, Nanjing 210023, People's Republic of China\\
$^{23}$Space Telescope Science Institute, 3700 San Martin Dr., Baltimore, MD 21218, USA\\
$^{24}$Department of Astronomy, The University of Texas at Austin, 2515
Speedway Blvd Stop C1400, Austin, TX 78712, USA\\
$^{25}$Department of Astronomy and Astrophysics, University of California, Santa Cruz, 1156 High Street, Santa Cruz, CA 95064, USA\\
$^{26}$Center for Computational Astrophysics, Flatiron Institute, 162 Fifth Avenue, New York, NY 10010, USA\\
$^{27}$Institute of Physics, GalSpec, Ecole Polytechnique Federale de Lausanne, Observatoire de Sauverny, Chemin Pegasi 51, 1290 Versoix, Switzerland\\
$^{28}$INAF, Astronomical Observatory of Trieste, Via Tiepolo 11, 34131 Trieste, Italy\\
$^{29}$Department of Computer Science, Aalto University, P.O. Box 15400, FI-00076 Espoo, Finland\\
$^{30}$Department of Physics, University of Helsinki, P.O. Box 64, FI-00014 Helsinki, Finland\\
$^{31}$SOFIA Science Center, NASA Ames Research Center, Moffett Field CA 94035, USA\\
$^{32}$Department of Physics \& Astronomy, University of California Los Angeles, 430 Portola Plaza, Los Angeles, CA 90095, USA
}


\bsp	
\label{lastpage}
\end{document}